\documentclass[twocolumn, a4paper, 11pt,book]{archive}
\usepackage{lineno,hyperref}
\modulolinenumbers[5]
\modulolinenumbers[1]
\usepackage{color}
\usepackage{multirow,bigdelim}
\usepackage{version}
\usepackage{graphicx}

\usepackage[]{natbib}

\usepackage{txfonts}

\usepackage{booktabs}
\usepackage{multirow}

\usepackage{rotating} 
\usepackage{color, colortbl}

\usepackage{siunitx}

\definecolor{Gray}{gray}{0.9}
\definecolor{LightRed}{rgb}{1., 0.88, 0.88}
\definecolor{LightCyan}{rgb}{0.88,1,1}
\definecolor{LightGreen}{rgb}{0.8,0.98,0.8}
\definecolor{LightViolet}{rgb}{0.95, 0.85, 1.0}
\definecolor{azur}{rgb}{0.94, 1.0, 1.0}

\begin{document} 

\twocolumn[{%
 \centering
%
{\center \bf \Huge Ethane clathrate hydrate infrared signatures for solar system remote sensing.}\\
\vspace*{0.25cm}

{\Large E. Dartois \inst{1},
        F. Langlet\inst{2},
 }\\
\vspace*{0.25cm}

$^1$      Institut des Sciences Mol\'eculaires d'Orsay, CNRS, Universit\'e Paris-Saclay, 
B\^at 520, Rue Andr\'e Rivi\`ere, 91405 Orsay, France\\
              \email{emmanuel.dartois@universite-paris-saclay.fr}\\
$^2$      Institut d'Astrophysique Spatiale (IAS), UMR8617, CNRS, Universit\'e Paris-Saclay, B\^at. 121, 91405 Orsay, France\\
 \vspace*{0.5cm}
{keywords: Planetary ices, IR spectroscopy, Clathrate hydrate, Ethane, Remote sensing}\\
 \vspace*{0.5cm}
{\it \large To appear in Icarus}\\
%
%
 }]
%
  \section*{Abstract}
  {Hydrocarbons such as methane and ethane are present in many solar system
objects, including comets, moons and planets. The interaction of these hydrocarbons with water ice at low temperatures could lead to the formation of inclusion compounds, such as clathrate hydrates (water based host cages trapping hydrocarbons guest molecules), modifying their retention, stability and therefore evolution. The occurrence of clathrate hydrates on solar system surfaces could be established by remote sensing of their spectroscopic signatures. In this study, we measure and analyse ethane clathrate hydrate spectra recorded in the temperature range from 5.3 to 160K, covering most of the temperature range of interest for solar system objects. Specific infrared band signatures are identified for the ethane encaged guest. 
We provide evidence that ethane clathrate hydrate outcrops can be detected by remote sensing on the surface of planetary bodies.
}
%

%

\section{Introduction}

Hydrocarbons are present in many solar system objects.
The most abundant hydrocarbons species include gaseous methane or ethane present in the atmosphere of giant planets \citep[e.g.][]{Melin2020, Guerlet2009, Sada1996, Tokunaga1975}, satellites atmospheres \citep{Cours2020, Lombardo2019, Niemann2010}, comets \citep{Villanueva2011, Crovisier2004, Mumma2001, Mumma2000}, in the liquid phase on Titan \citep{Farnsworth2019, Clark2010, Cordier2009, Lunine2008}, solid phase on Pluto and Triton \citep{DeMeo2010}, or Kuiper Belt objects \citep{Brown2007}.
The coexistence of hydrocarbons with water ice at low temperatures in some of these environments raises the question about their ability to form clathrate hydrates.
Clathrate hydrates are made of a 3-dimensional water-molecule connected network, linked via hydrogen bonds. To be stabilized, the clathrate network hosts cavities encapsulating guest molecules ("clathrate" meaning closure). Small molecules can be trapped in two main crystalline clathrate hydrate cubic structures named type I and type II. 
The type I structure's unit cell \citep[e.g.][]{Sloan2007} is made of two dodecahedral ($5^{12}$) water cages and six larger cages containing twelve pentagonal and two hexagonal faces  ($5^{12}6^{2}$) per unit cell. Another structure, the type II structure, possesses sixteen
 dodecahedral ($5^{12}$) cages for eight large cages with twelve pentagonal and four hexagonal faces ($5^{12}6^{4}$) per unit cell. For each structure, the cage's number ratio and the filling ratio with guest molecules influence the spectroscopic signatures of the trapped molecules.

When the guests are relatively volatile species, like hydrocarbons such as methane and ethane, clathrate hydrates will modify their retention time scales in solar system bodies, and also modify their release conditions. The hydrate term is sometimes used as a semantic shortcut to designate clathrate hydrates but must not be confused neither with hydrates nor with ice mixtures where a molecule is interacting with water ice and not necessarily mediated via an ordered crystalline cage.

Many models do include clathrate hydrates in the evolution or geophysical structure of solar system bodies, 
\citep[e.g.][]{Combe2019, Marounina2018, Castillo-Rogez2018, Fu2017, Luspay-Kuti2016, Marboeuf2012}.
A better understanding of clathrate kinetics and their spectroscopic signatures is needed to continue addressing constraints on these clathrate hydrate models.
A prerequisite for the identification of clathrate hydrates in the solar system is a thorough characterisation of the vibrational spectroscopic signatures in the infrared, or the specific spectral fingerprints, and their evolution with the ice structure and temperature. For small guest molecules, clathrate hydrates can crystallise in two different cubic structures (I and II), that will affect the filling of the ice cages and therefore the interactions and spectroscopy of the trapped species.
Methane is the main hydrocarbon found in cold solar system environments \citep[e.g.][and citations above]{Glein2020,Guzman2015}.
Dedicated experiments \citep[][and references therein]{Dartois2010, Dartois2009, Dartois2008} have recorded the clathrate hydrate infrared signatures in the 4K to 140K range, in the 2-4~$\mu$m (5000-2500 cm$^{-1}$) range covering the fundamental stretching modes as well as the strongest combinations/overtones.

In addition to methane, ethane has been detected on Titan and contributes to the hydrocarbon cycle \citep{Brown2008, Mousis2008, Barth2006, Griffith2006, Lunine1983}.
In some instances, such as the surface of Titan, the lower abundance of the ethane hydrocarbon with respect to methane can 
be compensated by ethane's lower vapour pressure, or its liquid state, and could favour the ethane inclusion in the formation of clathrate hydrates. 
Missions targeting Titan, such as Dragonfly should better constrain the surface and lower atmosphere conditions \citep[e.g.][]{Turtle2020, Voosen2019, Lorenz2018}.
In this article we explore the infrared spectra of the ethane clathrate hydrate in the 5.3-160K range, and provide the spectroscopic signature and band frequencies of the guest ethane trapped molecules.
%
\begin{figure*}[htbp]
\centering
\includegraphics[angle=0,width=\columnwidth]{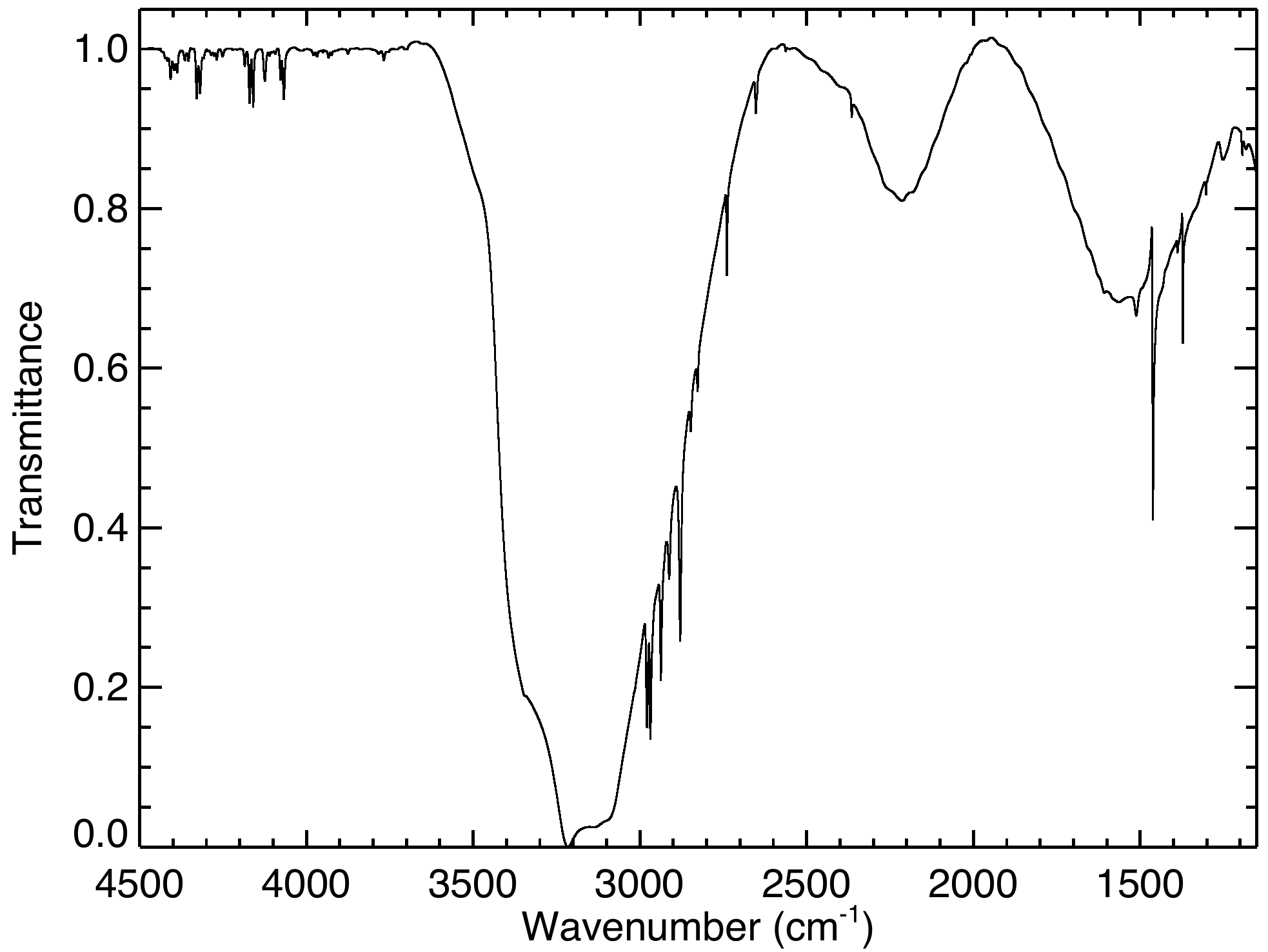}
\caption{Baseline corrected ethane clathrate hydrate transmittance spectrum recorded at 5.3K. Note the imperfect cancelation of the Fabry-Perot fringes in the space between the ZnSe windows, leading to small waves observed superimposed on the spectrum (e.g. in the 2500-1700 cm$^{-1}$ interval). Scattering effects deform the water ice bands because of imperfect film surface roughness once clathration is achieved.}
\label{Fig_generale}
\end{figure*}
%
\begin{figure*}[htbp]
\centering
\includegraphics[angle=0,width=\columnwidth]{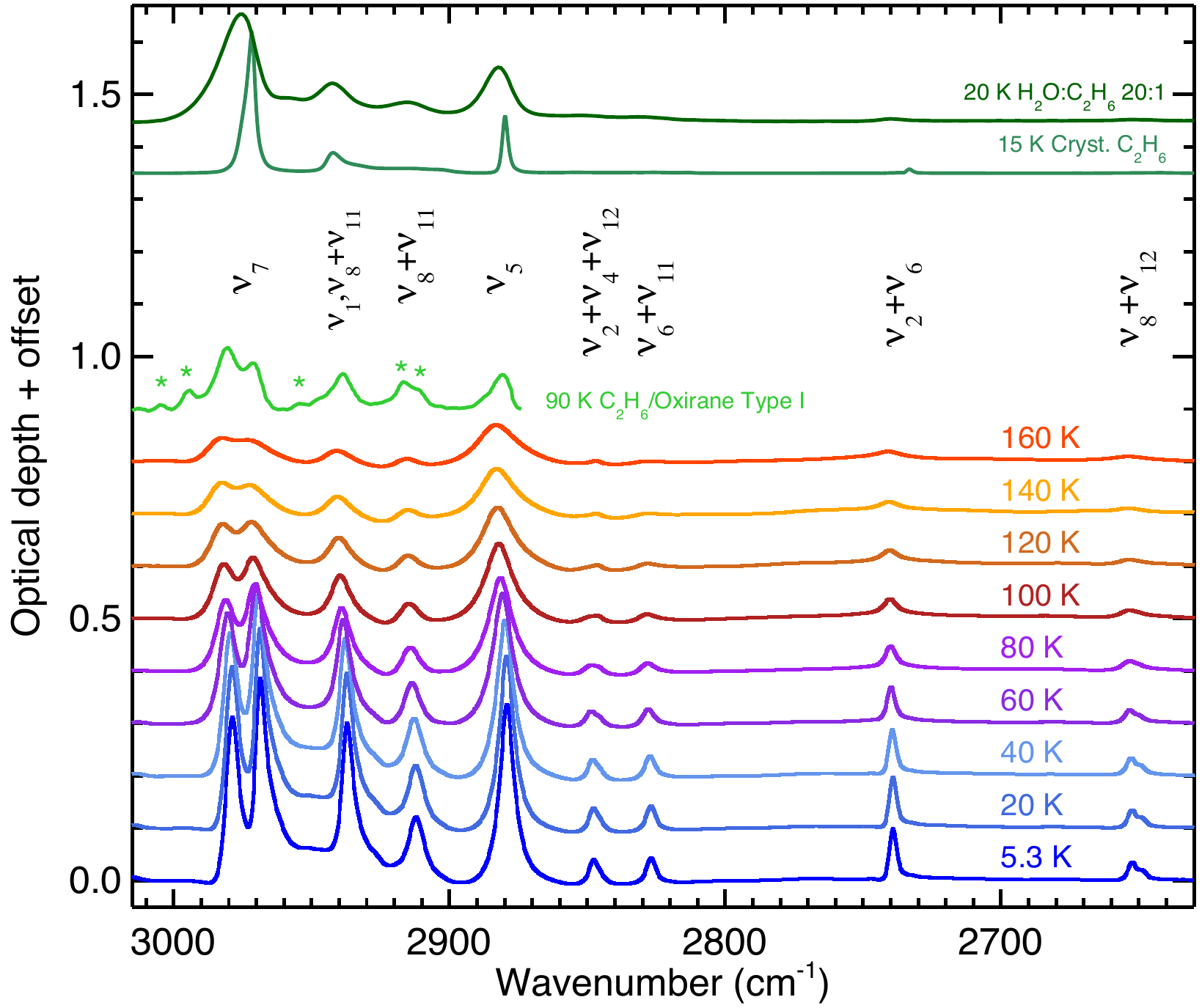}
\includegraphics[angle=0,width=\columnwidth]{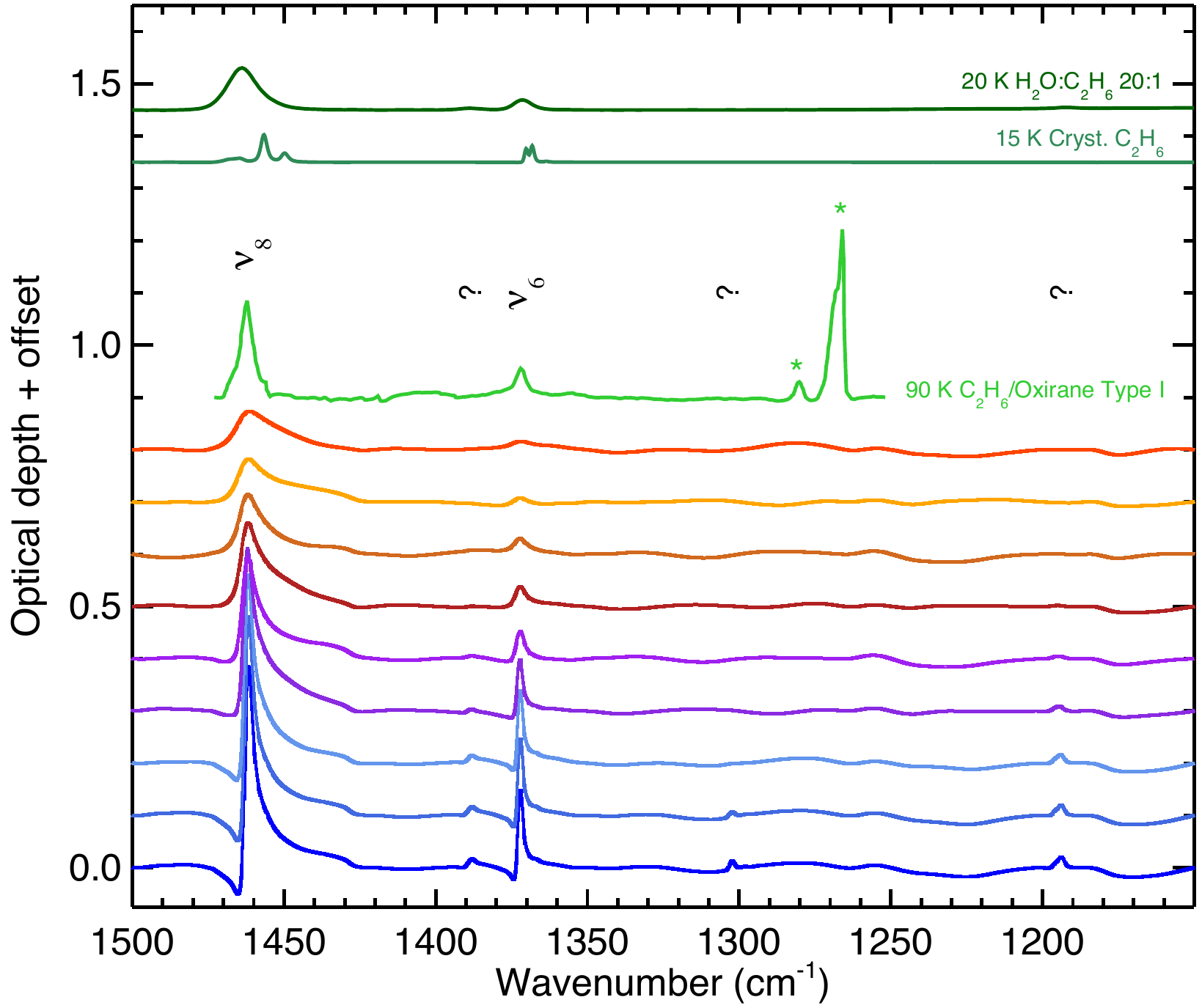}
\caption{Temperature-dependent infrared spectra of the ethane clathrate in the fundamental modes regions. The crystalline pure ethane spectrum of \cite{Hudson2014} recorded at 15K is shown above. The infrared spectrum of an H$_2$O:C$_2$H$_6$ (20:1) ice mixture from the Cosmic Ice Laboratory (Hudson et al., https://science.gsfc.nasa.gov/691/cosmicice/spectra.html), 
recorded at 20K, is shown above. The ethane:oxirane (7:3) mixed clathrate spectrum from \cite{Richardson1985} recorded at 90K is also shown. It is baseline corrected, and oxirane features are labelled with asterisks for clarity. Tentative assignments of the implied vibrational modes are given (see text for details). Spectra are offset for clarity.}
\label{Fig_clathrate_fundamental}
\end{figure*}

\begin{figure*}[htbp]
\centering
\includegraphics[angle=0,width=\columnwidth]{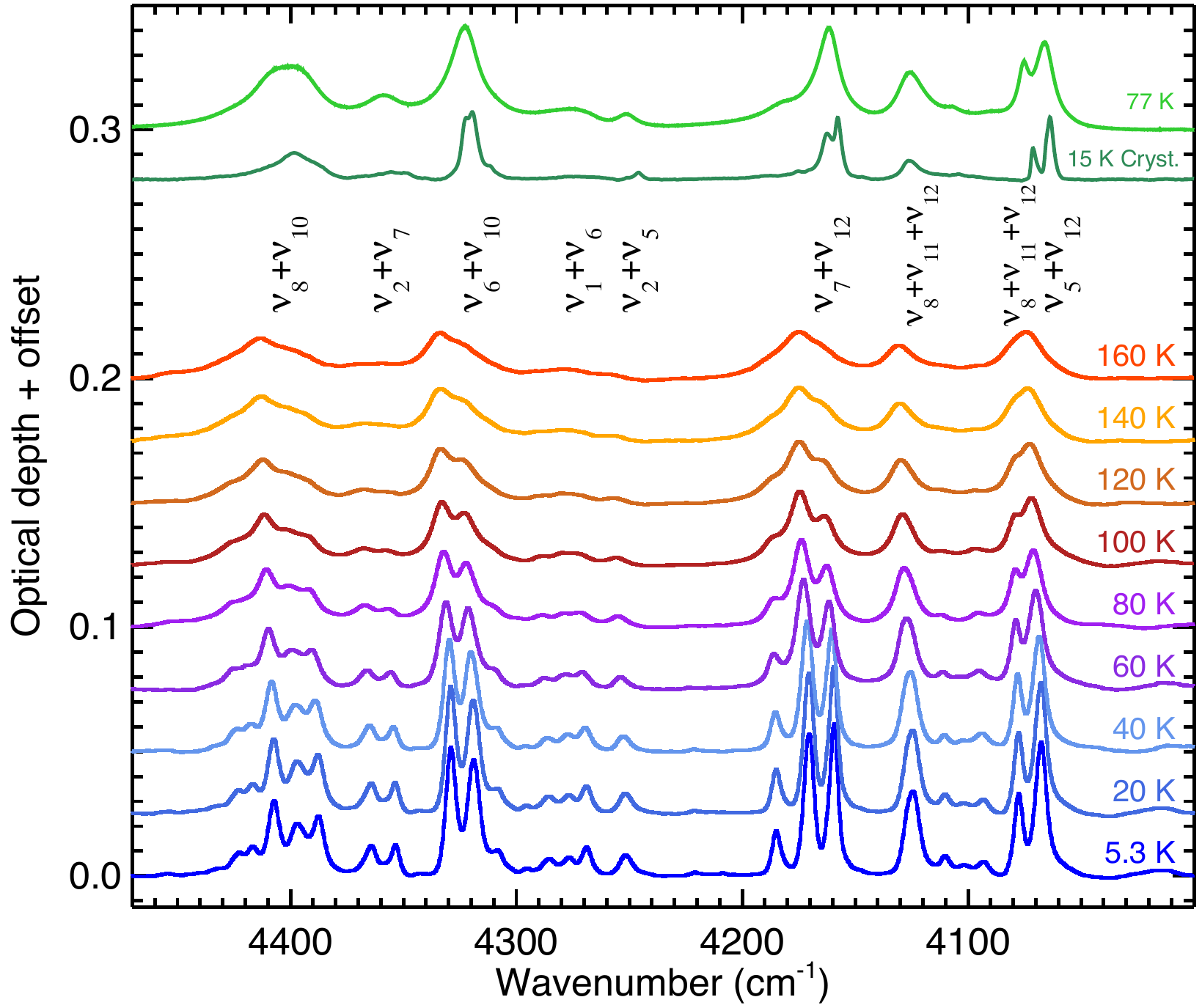}
\includegraphics[angle=0,width=\columnwidth]{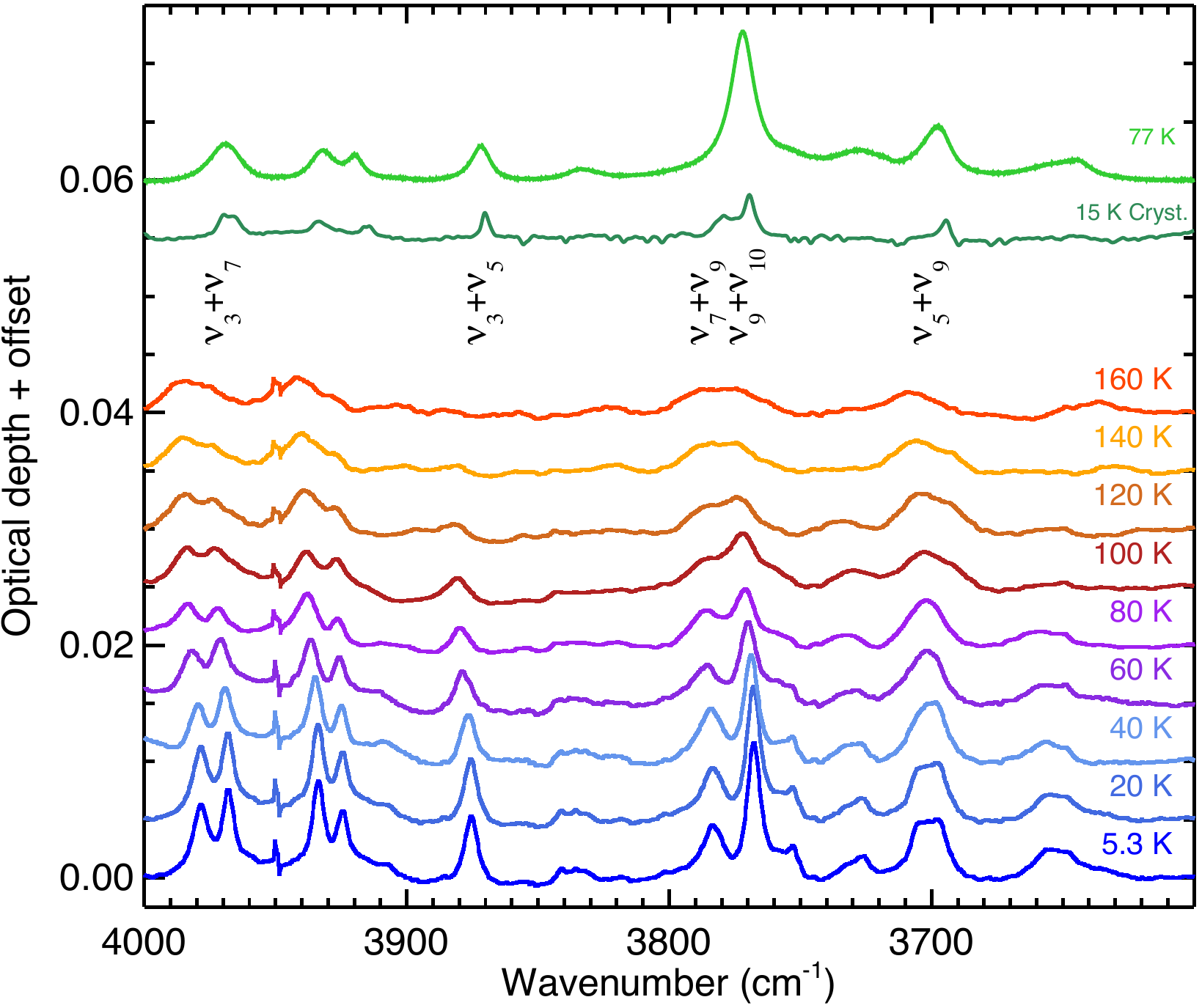}
\caption{Temperature dependent infrared spectra of the ethane clathrate in the combination/overtone regions. The crystalline pure ethane spectrum of \cite{Hudson2014} recorded at 15K is shown above. The pure ethane spectrum recorded in the cell at 77K is shown above too. Tentative assignments of the implied vibrational modes are given (see text for details). Spectra are offset for clarity.}
\label{Fig_clathrate_combinations}
\end{figure*}

\begin{figure*}[htbp]
\centering
\includegraphics[angle=0,width=2\columnwidth]{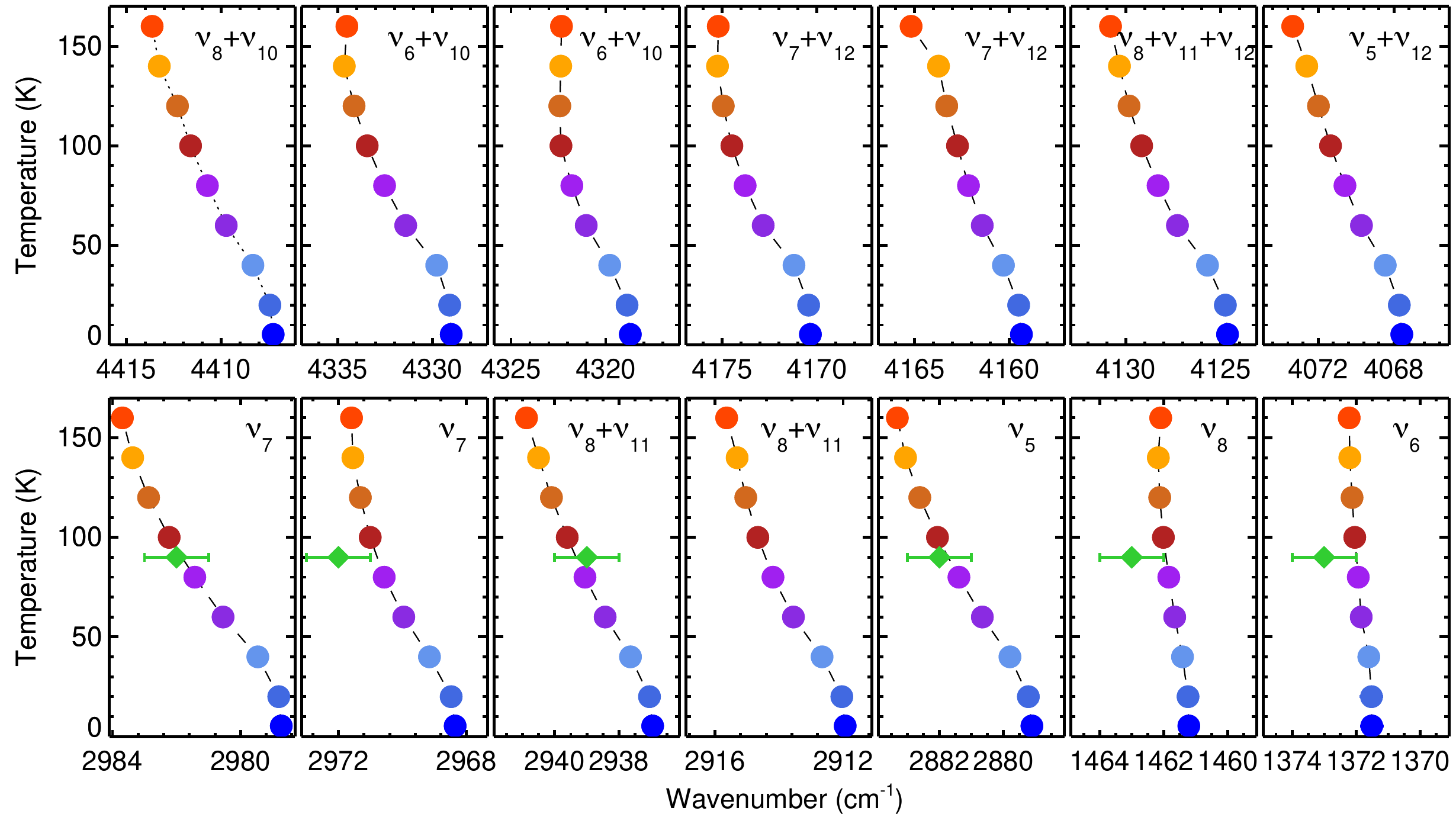}
\caption{Temperature dependent positions for selected transitions (circles), with their tentative vibrational assignments, given in Table~\ref{positions}. The ethane centroid vibrational positions for an ethane:oxirane (7:3) mixed clathrate spectrum from \cite{Richardson1985} recorded at 90K, are shown with green diamonds. Oxirane is known to form and induce a clathrate hydrate type I structure, helping to confirm the structure formed and also assign the C$_2$H$_6$ vibrational bands in this mixed oxirane clathrate hydrate.}
\label{Fig_clathrate_positions}
\end{figure*}

%
\section{Experiments}
To form the ethane clathrate hydrate we followed a previously proven protocol in evacuated cryogenic cells \citep{Dartois2010, Dartois2009, Dartois2008}. The cell was built around two ZnSe windows facing each other, for infrared transmission analysis, sealed to an oxygen-free high thermal conductivity (OFHC), gold coated, copper closed cell. It was thermally coupled to a cold finger whose temperature can be lowered using a liquid He transfer, balanced by a surrounding Minco polyimide thermofoil heater, which maintains a constant temperature.
A high vacuum evacuated cryostat ($\rm P < 10^{-7}$mbar) surrounded the cell. Gases were injected into the cell (or evacuated) with a stainless steel injection tube brazed at the bottom of the cell.
In these experiments, a water ice film was formed first, by injecting water vapour into the cell maintained just below the water ice freezing point, forming an Ih ice. 
De-ionized water was used. Several freeze-pump-thaw degassing cycles were applied to remove any dissolved gases.
The cell temperature was then lowered to 267~K for the clathrate formation. The cell was pressurised with ethane gas at about 25 bars. This \{pressure,temperature\} couple is well above the stability curve for ethane clathrate hydrate, which is about 3.5 bars at 267~K. The system was kept under pressure for approximately fifteen days, much above the expected diffusion kinetics time scale. If one assumes a typical diffusion coefficient around $\rm10^{-12}m^2/h$ at 267K \citep[e.g.][and references therein]{Falenty2013}, in fifteen days the diffusion pervaded $\sim$20$\mu$m. The deposited water ice film thickness was several microns thick (the film deposited on each face of the cell is about 3$\pm1$$\mu$m thick), thus fully penetrated by diffusion by the ethane gas to make the clathrate.
The temperature was then progressively lowered (at a rate of $\sim$1K/min) while evacuating the ethane gas.
During this operation, we maintained the system just below the sublimation/vaporisation curve of pure ethane, and above the ethane clathrate hydrate expected stability curve \citep{Roberts1940}. 

When the system was at a temperature sufficiently low enough so that the kinetics of the clathrate destabilisation becomes very long (typically below 100~K), the system was fully evacuated while the temperature continued to be lowered to its minimum of about 5.3K.
 The infrared spectra were then recorded with steps of 20K, which allowed us to follow the evolution of the spectral signatures, up to the clathrate hydrate rapid dissociation occurring above 160~K over an hour under our experimental conditions. 
 (In between each temperature stabilised step, the temperature was raised at a rate of $\sim$1K/min, allowing for proper thermalisation of the clathrate hydrate.)
The spectra were recorded with a Bruker Fourier transform infrared spectrometer covering the 5000-700 cm$^{-1}$ spectral range at a resolution of 0.5~cm$^{-1}$, with a globar source, KBr beamsplitter and an HgCdTe detector cooled with liquid N$_2$. 
\begin{table*}[htp]
\caption{Observed clathrate hydrate ethane positions and tentative assignments}
\begin{center}
\begin{tabular}{lllll}
\hline
\multicolumn{2}{c}{$\rm \bar{\nu}_{CLH}$ (cm$^{-1}$)$^a$}  	& \multicolumn{2}{c}{Tentative assignments$^b$}	&$\rm \bar{\nu}_{gas}$ (cm$^{-1}$)$^b$\\
\hline
160~K				&5.3~K	&		&	\\
\hline
					&4423.7	$\pm$0.3		& 				&	\\
					&4416.7	$\pm$0.3		& 				&	\\
4413.8 $\pm$0.3		&4407.3			  	&\rdelim\}{3}{5pt}[] 	&\multirow{3}{*}{$\nu_8+\nu_{10}$}	&\multirow{3}{*}{4414.44}	\\
					&4397.0 				& 				&	\\
					&4387.3				& 				&	\\
					&4364.4				&\rdelim\}{2}{5pt}[] 	&\multirow{2}{*}{$\nu_2+\nu_7$}	&\multirow{2}{*}{4377.67}\\
					&4353.4				& 				&	\\
4334.5 $\pm$0.3		&4329.1				&\rdelim\}{3}{5pt}[]  	&\multirow{3}{*}{$\nu_6+\nu_{10}$}	&\multirow{3}{*}{4342.56}\\
4322.6 $\pm$0.3		&4318.7				& 				&	\\
					&4307.7				& 				&	\\
					&4295.6	$\pm$0.3		& 				&	\\
					&4285.7				&\rdelim\}{3}{5pt}[] 	&\multirow{3}{*}{$\nu_1+\nu_{6}$}	&\multirow{3}{*}{4293.40}\\
					&4276.8				& 				&	\\
					&4268.7				& 				&	\\
					&4251.6				& 				&$\nu_2+\nu_{5}$	&4274.633\\
					&4184.9	 			& 				&	\\
4175.2 $\pm$0.3		&4170.4				&\rdelim\}{2}{5pt}[]  	&\multirow{2}{*}{$\nu_7+\nu_{12}$}	&\multirow{2}{*}{4179.835, 4177.523}\\
4165.5 $\pm$1.0		&4159.4				& 				&	\\
4130.9 $\pm$0.3		&4124.8	$\pm$0.3		& 				&$\nu_8+\nu_{11}+\nu_{12}$		&4145.67\\
					&4110.1				& 				&	\\
					&4101.6	$\pm$0.5		& 				&	\\
					&4092.9				& 				&	\\
					&4077.7				& 				&$\nu_8+\nu_{11}+\nu_{12}$		&4098.2209\\
4074.5 $\pm$0.3		&4067.6				& 				&$\nu_5+\nu_{12}$				&4086.20\\
3985.5 $\pm$1.0		&3978.6   				& \rdelim\}{2}{5pt}[]	&\multirow{2}{*}{$\nu_3+\nu_7$}	&\multirow{2}{*}{\it(994.11+2985.39)$^c$}\\
					&3967.9   				& 				&	\\
3942.0 $\pm$1.0		&3933.7   				& 				&	\\
					&3924.3   				& 				&	\\
					&3875.4   				& 				&$\nu_3+\nu_5$				&\it(994.11+2895.67)$^c$\\
					&3783.0  $\pm$0.3		& 				&$\nu_7+\nu_9$				&\it(2985.39+821.72)$^c$\\
3774.0 $\pm$2.0		&3767.7   				& 				&$\nu_9+\nu_{10}$				&\it(821.72+2968.69)$^c$\\
					&3752.4  $\pm$0.5		& 				&	\\
					&3726.2  				& 				&	\\
3710.0 $\pm$2.0		&3700.2$^d$			& 				&$\nu_5+\nu_{9}$				&\it(2895.67+821.72)$^c$\\
					&3656.5$^d$			& 				&	\\
					&3648.0$^d$			& 				&	\\
2983.7 $\pm$0.3		&2978.8   				&\multirow{2}{*}{\}} 	&\multirow{2}{*}{$\nu_7$}			&\multirow{2}{*}{2985.39}\\
2969.8 $\pm$0.8		&2968.3    			& 				&	\\
2940.8 $\pm$0.3		&2937.0    			& 				&$\nu_8+\nu_{11},\nu_1$			&2953.8, 2954.0\\
2914.8 $\pm$0.8		&2911.9    			& 				&$\nu_8+\nu_{11}$				&2930.705\\
2883.3 $\pm$0.3		&2879.1    			& 				&$\nu_5$						&2895.67\\
2846.9$^d$ $\pm$0.6	&2847.5$^d$    			& 				&$\nu_2+\nu_4+\nu_{12}$			&2860	\\
2827.6 $\pm$0.4		&2826.8    			& 				&$\nu_6+\nu_{11}$				&2844.13069\\
2741.0 $\pm$0.3		&2739.1   				& 				&$\nu_2+\nu_6$				&2753.326\\
2653.2$^d$ $\pm$0.5	&2652.4   				&\multirow{2}{*}{\}} 	&\multirow{2}{*}{$\nu_8+\nu_{12}$}	&\multirow{2}{*}{2665.1512}	\\
					&2648.9   				& 				&	\\
2367.0 $\pm$0.5		&2365.0  				& 				&$\nu_3+\nu_{6}$				&2368.59\\
1462.2 $\pm$0.5		&1461.6   				& 				&$\nu_8$						&1472.03	\\
					&1388.0				& 				&?				&	\\
1372.2 $\pm$0.5		&1372.1				& 				&$\nu_6$						&1379.163\\
					&1302.4				& 				&?				&	\\
					&1194.1				& 				&?				&	\\
\hline
\end{tabular}
\end{center}
\label{positions}
$^a$ At 5.3K, unless stated, positions uncertainties for the strongest bands are of the order of $\pm$0.2~cm$^{-1}$. 
$^b$ Tentative assignments are based on gas phase positions from \cite{Hepp2000}, and references therein, assuming a redshift induced by the clathrate encaging.
$^c$ estimated assignment from the sum of fundamental modes positions.
$^d$ broad, possibly double.
\end{table*}
\section{Results}
The ethane clathrate hydrate transmittance spectrum, at 5.3K, after baseline correction, is shown in Fig.\ref{Fig_generale}. The ethane clathrate hydrate bands in the infrared recorded from 5.3 to 160K are shown in Fig.\ref{Fig_clathrate_fundamental} for the fundamental vibrational modes range (3000-2700 and 1500-1100 cm$^{-1}$), and in Fig.\ref{Fig_clathrate_combinations} for the combinations/overtones (4470-3600 cm$^{-1}$). These spectra are compared to the spectra of pure ethane ice recorded in the same cell at 77K, crystalline ethane recorded at 15K \citep{Hudson2014}, and an amorphous H$_2$O:C$_2$H$_6$ (20:1) ice mixture from the Cosmic Ice Laboratory\footnote{Hudson et al., https://science.gsfc.nasa.gov/691/cosmicice/spectra.html} in the fundamental modes regions.
A spectrum of an ethane:oxirane (7:3) mixed clathrate from \cite{Richardson1985} recorded at 90K, is shown above the spectra in the mid-infrared. The oxirane vibrational mode contribution is identified with asterisks. 
Oxirane is known to form and induce a clathrate hydrate type I structure. Its clathrate type I and II spectral positions are well known, helping to confirm the structure formed and thus assign the vibrational spectrum for other guests trapped in mixed oxirane clathrate hydrates.
A tentative attribution of the vibrational modes and combinations/overtones is over-plotted on these figures. The assignments are based on the gas phase rovibrational analysis presented by \cite{Hepp2000,Hepp1999,Hepp1999b,Hepp1998,Hepp1997}, and assumes a bathochromic shift (a few wavenumbers for fundamental vibrations and up to ten to twenty for the combination modes regions) induced by the ethane clathration in a clathrate hydrate large cage (see Table~\ref{positions}).
We also report the band position evolution as a function of temperature in Fig.\ref{Fig_clathrate_positions} for selected intense transitions, %
and compare these band transitions with positions from the ethane:oxirane mixed clathrate spectrum from \cite{Richardson1985} for the mid-infrared modes.
\section{Discussion}
Pure ethane clathrate hydrate is expected to form a type I clathrate hydrate. 
The positions of the ethane bands in our experiment, when compared to the \cite{Richardson1985} mixed ethane:oxirane band positions (shown with green diamond in Fig.\ref{Fig_clathrate_positions}) coincide well with the expected type I clathrate hydrate positions. \cite{Richardson1985} however point toward unusual relative intensities of the oxirane band intensities in this experiment, requiring further analysis. It may suggest that the presence of ethane in this mixed clathrate may deform the structure more than for smaller guest species.
Peak positions and tentative assignments are reported in Table~\ref{positions} for the two measured extreme temperatures, 5.3 and 160~K, respectively.\\

The main outcome of this study is that the formation of an ethane clathrate hydrate gives rise to a recorded very unique band pattern, distinct from the other observed features for ethane in other possible phases.
In particular the band splitting in the fundamental $\nu_7$ degenerate mode region around 2970~cm$^{-1}$ (already observed by \cite{Richardson1985}), with a splitting amplitude of a bit more than 10~cm$^{-1}$) is characteristic of the ethane clathrate hydrate, and not observed in other pure ethane or mixed ethane/water ice phases. 
In the combinations/overtone near infrared region, many bands observed in the pattern are also characteristic (in particular the $\sim$4400, 4320, 4170, 3980 cm$^{-1}$ combinations modes splittings). They are clearly distinctive from the other possible solid ethane or ethane simple hydrate phases, pointing toward the uniqueness of the identification of a clathrate hydrate through the observation of these modes.\\

In planetary environments such as Jupiter, Saturn and its satellite Titan, Pluto and other TNOs, as well as comets, ethane is less abundant than methane.
When type I clathrate hydrate formers are mixed, they may induce the formation of a type II hydrate, under specific ratios. \cite{Subramanian2000} report that in mixed methane and ethane clathrate hydrate studied by Raman spectroscopy, both type I formers taken individually, formed a type II clathrate hydrate under specific composition ranges, typically starting above 72 to 75\% of methane. The structure will go back to type I clathrate hydrate when ethane becomes a trace species, above the upper bound of 95\% methane in these experiments. 
The mixed methane/ethane clathrate hydrate is metastable in some cases as shown by, e.g., \citet{Takeya2003}.
\cite{Takeya2005} measured the lattice constants of mixed methane and ethane clathrate hydrate at 113 K as a function of CH$_4$ gas content. The mixed system forms the type II clathrate hydrate for about 60 to 80\% methane content and type I above and below, with values close to the pure methane hydrate lattice parameters of type I at high methane content.
Type I and type II mixed methane+ethane clathrate hydrates can coexist. The type II has been shown to be a metastable kinetically favoured stage during some mixed clathrate hydrate nucleation experiments, with a slow transformation into sI, the type I being the stable phase to long time scales
\citep{Kida2010,Ohno2009,Murshed2009}.
In both type I and II clathrates, the ethane molecules occupies the large cage (the $5^{12}6^{2}$ in type I and $5^{12}6^{4}$ in type II), as ethane is too large to fit into the small cage. The proportions at which this change occurs will vary with the formation temperature of the clathrate hydrate because of the difference in fugacity of these two species.
The ratio of small to large cages in each unit of the two clathrate ice structures favours type I when only the large cages are filled. The band positions are expected to shift by a few wavenumbers in between type I and type II clathrate hydrate large cages, being about 4 cm$^{-1}$ redder for type II with respect to type I in the stretching modes, based on Raman spectroscopy \citep{Subramanian2000}. 
Apart from the shift specific to each clathrate hydrate structure, the pattern observed in the present experiments should be representative. 
%
Several splittings for some vibrational modes/combinations, distinct from the known crystalline phase splittings, characteristic of ethane trapped in the large cages of a clathrate hydrate of type I (or II with an expected moderate shift) should be actively searched for to establish such clathrate hydrates occurence.\\

Saturn's moon Titan, with an atmospheric temperature in the 70-180K range \citep{Horst2017, Niemann2010, McKay1989}, hosts conditions for hydrocarbon rains, and, with a surface temperature around 90K, harbours liquid hydrocarbons, including methane and ethane lakes \citep[e.g.][]{Mastrogiuseppe2019, Hayes2008, Stofan2007}. The stability zone for the formation of methane and ethane clathrate hydrate falls in this temperature range \citep{Kalousova2020, Tobie2009, Lunine2008, Thomas2008, Thomas2007, Atreya2006, Osegovic2005}. 
\cite{Vu2020} recently showed that the kinetics of formation at low temperature of an ethane clathrate hydrate is kinetically possible under Titan conditions, using Raman spectroscopy. 
These authors report that with a surface temperature of 90K, a full clathration of ice in contact with ethane will occur in about four (Earth) years, i.e. within a time scale shorter than Titan's seasons (winter and summer solstices are separated by about 15 years, directly related to Saturn's orbital period of slightly more than 29 years). An alternative formation at the surface invoked in this study is the contact of ethane rain directly on exposed ice slabs, a situation for which the clathrate hydrate kinetics is more difficult to estimate. However, for such exposed fields, the spectroscopic signatures obtained by remote sensing are directly accessible, and would provide constraints on clathrates occurrence. In Titan's lakes, the density of liquid ethane being higher than liquid methane (e.g., about 462.7 kg/m$^3$ (melting line) and 662.8 kg/m$^3$ (melting line) at 100K for methane and ethane, respectively \citep{Younglove1987}), a methane/ethane phase separation should occur. The full nitrogen, methane and ethane system must be investigated under Titan conditions \citep[e.g.][]{Hanley2017}. At the bottom of these lakes, 
ethane may therefore be in direct contact with an ice bedrock like the one observed elsewhere on Titan \citep[e.g.][]{Griffith2019}, but residing well below the lake's surface.
Additionally, if the temperature at the bottom of the lakes is higher than the surface temperature, the formation kinetics should be thermally enhanced.
\cite{Vu2020} consider the domain of formation at greater depths if ethane percolates deeper through the surface, with corresponding higher temperatures and favourable formation kinetics, but hardly accessible to spectroscopic means.
Many experiments following the cage evolution and spectral signatures of clathrate hydrates are performed using Raman spectroscopy. In the spectral range considered here, if both Raman and infrared spectrometers can be integrated in space probes that would land on solar system objects, and perform an analysis of the local field, a vast majority of the spectroscopic information is still obtained using infrared only remote sensing by space probes using on-board spectrometers.
An outcrop on the surface of icy bodies, or a terrain recently exposed when there is a space probe fly-by may allow for the identification of clathrate hydrates by infrared spectroscopic means.
Such a direct spectroscopic detection would considerably anchor our view on the geophysical evolution of these objects.
\section{Conclusion}
We measured the temperature dependence, in the 5.3 to 160~K range, of the near and mid-infrared spectral signatures of ethane trapped in a type I clathrate hydrate structure, identified by comparison to literature spectra of mixed oxirane:ethane clathrate hydrate. 
The ethane molecule occupies the large cages of the ice crystal structure formed by the host water molecules. 
The large number of individual transitions observed in the fundamental modes of vibration and the numerous combinations/harmonics represent distinctive spectroscopic fingerprints that will allow remote identification if ethane is present in a clathrate hydrate in solar system ices. 
Further experiments are foreseen to explore the type I and II clathrate hydrate differences in the infrared over a large range of temperatures. 
When ethane occupies the large cage of type I or II clathrate hydrates, only a mild shift of about four wavenumbers between ethane bands in each structure is expected, and, apart from a shift of such magnitude, the measured spectra in this study will be similar to those of both structures.
These signatures can be used to analyse planetary remote sensing data and may be used for future proposed missions exploring the icy worlds of the solar system, in particular, ones targeting Saturn's moon Titan, such as the Dragonfly mission.
\section*{Acknowledgments}
This work was supported by the CNRS/INSU and CNRS/INP. The authors want to thank M. Bouzit, B. Crane, for their technical support to the initial design of the cell.
%
%
\bibliographystyle{model2-names}

\bibliography{clathrate_c2h6.bib}

\begin{thebibliography}{67}
\expandafter\ifx\csname natexlab\endcsname\relax\def\natexlab#1{#1}\fi
\providecommand{\url}[1]{\texttt{#1}}
\providecommand{\href}[2]{#2}
\providecommand{\path}[1]{#1}
\providecommand{\DOIprefix}{doi:}
\providecommand{\ArXivprefix}{arXiv:}
\providecommand{\URLprefix}{URL: }
\providecommand{\Pubmedprefix}{pmid:}
\providecommand{\doi}[1]{\href{http://dx.doi.org/#1}{\path{#1}}}
\providecommand{\Pubmed}[1]{\href{pmid:#1}{\path{#1}}}
\providecommand{\bibinfo}[2]{#2}
\ifx\xfnm\relax \def\xfnm[#1]{\unskip,\space#1}\fi
\bibitem[{{Atreya} et~al.(2006){Atreya}, {Adams}, {Niemann},
  {Demick-Montelara}, {Owen}, {Fulchignoni}, {Ferri} and {Wilson}}]{Atreya2006}
\bibinfo{author}{{Atreya}, S.K.}, \bibinfo{author}{{Adams}, E.Y.},
  \bibinfo{author}{{Niemann}, H.B.}, \bibinfo{author}{{Demick-Montelara},
  J.E.}, \bibinfo{author}{{Owen}, T.C.}, \bibinfo{author}{{Fulchignoni}, M.},
  \bibinfo{author}{{Ferri}, F.}, \bibinfo{author}{{Wilson}, E.H.},
  \bibinfo{year}{2006}.
\newblock \bibinfo{title}{{Titan's methane cycle}}.
\newblock \bibinfo{journal}{Planetary and Space Science} \bibinfo{volume}{54},
  \bibinfo{pages}{1177--1187}.
\newblock \DOIprefix\doi{10.1016/j.pss.2006.05.028}.
\bibitem[{{Barth} and {Toon}(2006)}]{Barth2006}
\bibinfo{author}{{Barth}, E.L.}, \bibinfo{author}{{Toon}, O.B.},
  \bibinfo{year}{2006}.
\newblock \bibinfo{title}{{Methane, ethane, and mixed clouds in Titan's
  atmosphere: Properties derived from microphysical modeling}}.
\newblock \bibinfo{journal}{Icarus} \bibinfo{volume}{182},
  \bibinfo{pages}{230--250}.
\newblock \DOIprefix\doi{10.1016/j.icarus.2005.12.017}.
\bibitem[{{Brown} et~al.(2007){Brown}, {Barkume}, {Blake}, {Schaller},
  {Rabinowitz}, {Roe} and {Trujillo}}]{Brown2007}
\bibinfo{author}{{Brown}, M.E.}, \bibinfo{author}{{Barkume}, K.M.},
  \bibinfo{author}{{Blake}, G.A.}, \bibinfo{author}{{Schaller}, E.L.},
  \bibinfo{author}{{Rabinowitz}, D.L.}, \bibinfo{author}{{Roe}, H.G.},
  \bibinfo{author}{{Trujillo}, C.A.}, \bibinfo{year}{2007}.
\newblock \bibinfo{title}{{Methane and Ethane on the Bright Kuiper Belt Object
  2005 FY9}}.
\newblock \bibinfo{journal}{Astronomical Journal} \bibinfo{volume}{133},
  \bibinfo{pages}{284--289}.
\newblock \DOIprefix\doi{10.1086/509734}.
\bibitem[{{Brown} et~al.(2008){Brown}, {Soderblom}, {Soderblom}, {Clark},
  {Jaumann}, {Barnes}, {Sotin}, {Buratti}, {Baines} and
  {Nicholson}}]{Brown2008}
\bibinfo{author}{{Brown}, R.H.}, \bibinfo{author}{{Soderblom}, L.A.},
  \bibinfo{author}{{Soderblom}, J.M.}, \bibinfo{author}{{Clark}, R.N.},
  \bibinfo{author}{{Jaumann}, R.}, \bibinfo{author}{{Barnes}, J.W.},
  \bibinfo{author}{{Sotin}, C.}, \bibinfo{author}{{Buratti}, B.},
  \bibinfo{author}{{Baines}, K.H.}, \bibinfo{author}{{Nicholson}, P.D.},
  \bibinfo{year}{2008}.
\newblock \bibinfo{title}{{The identification of liquid ethane in Titan's
  Ontario Lacus}}.
\newblock \bibinfo{journal}{Nature} \bibinfo{volume}{454},
  \bibinfo{pages}{607--610}.
\newblock \DOIprefix\doi{10.1038/nature07100}.
\bibitem[{{Castillo-Rogez} et~al.(2018){Castillo-Rogez}, {Neveu}, {McSween},
  {Fu}, {Toplis} and {Prettyman}}]{Castillo-Rogez2018}
\bibinfo{author}{{Castillo-Rogez}, J.}, \bibinfo{author}{{Neveu}, M.},
  \bibinfo{author}{{McSween}, H.Y.}, \bibinfo{author}{{Fu}, R.R.},
  \bibinfo{author}{{Toplis}, M.J.}, \bibinfo{author}{{Prettyman}, T.},
  \bibinfo{year}{2018}.
\newblock \bibinfo{title}{{Insights into Ceres's evolution from surface
  composition}}.
\newblock \bibinfo{journal}{Meteoritics and Planetary Science}
  \bibinfo{volume}{53}, \bibinfo{pages}{1820--1843}.
\newblock \DOIprefix\doi{10.1111/maps.13181}.
\bibitem[{{Clark} et~al.(2010){Clark}, {Curchin}, {Barnes}, {Jaumann},
  {Soderblom}, {Cruikshank}, {Brown}, {Rodriguez}, {Lunine}, {Stephan},
  {Hoefen}, {Le Mou{\'e}lic}, {Sotin}, {Baines}, {Buratti} and
  {Nicholson}}]{Clark2010}
\bibinfo{author}{{Clark}, R.N.}, \bibinfo{author}{{Curchin}, J.M.},
  \bibinfo{author}{{Barnes}, J.W.}, \bibinfo{author}{{Jaumann}, R.},
  \bibinfo{author}{{Soderblom}, L.}, \bibinfo{author}{{Cruikshank}, D.P.},
  \bibinfo{author}{{Brown}, R.H.}, \bibinfo{author}{{Rodriguez}, S.},
  \bibinfo{author}{{Lunine}, J.}, \bibinfo{author}{{Stephan}, K.},
  \bibinfo{author}{{Hoefen}, T.M.}, \bibinfo{author}{{Le Mou{\'e}lic}, S.},
  \bibinfo{author}{{Sotin}, C.}, \bibinfo{author}{{Baines}, K.H.},
  \bibinfo{author}{{Buratti}, B.J.}, \bibinfo{author}{{Nicholson}, P.D.},
  \bibinfo{year}{2010}.
\newblock \bibinfo{title}{{Detection and mapping of hydrocarbon deposits on
  Titan}}.
\newblock \bibinfo{journal}{Journal of Geophysical Research (Planets)}
  \bibinfo{volume}{115}, \bibinfo{pages}{E10005}.
\newblock \DOIprefix\doi{10.1029/2009JE003369}.
\bibitem[{{Combe} et~al.(2019){Combe}, {McCord}, {Matson}, {Johnson}, {Davies},
  {Scipioni} and {Tosi}}]{Combe2019}
\bibinfo{author}{{Combe}, J.P.}, \bibinfo{author}{{McCord}, T.B.},
  \bibinfo{author}{{Matson}, D.L.}, \bibinfo{author}{{Johnson}, T.V.},
  \bibinfo{author}{{Davies}, A.G.}, \bibinfo{author}{{Scipioni}, F.},
  \bibinfo{author}{{Tosi}, F.}, \bibinfo{year}{2019}.
\newblock \bibinfo{title}{{Nature, distribution and origin of CO$_{2}$ on
  Enceladus}}.
\newblock \bibinfo{journal}{Icarus} \bibinfo{volume}{317},
  \bibinfo{pages}{491--508}.
\newblock \DOIprefix\doi{10.1016/j.icarus.2018.08.007}.
\bibitem[{{Cordier} et~al.(2009){Cordier}, {Mousis}, {Lunine}, {Lavvas} and
  {Vuitton}}]{Cordier2009}
\bibinfo{author}{{Cordier}, D.}, \bibinfo{author}{{Mousis}, O.},
  \bibinfo{author}{{Lunine}, J.I.}, \bibinfo{author}{{Lavvas}, P.},
  \bibinfo{author}{{Vuitton}, V.}, \bibinfo{year}{2009}.
\newblock \bibinfo{title}{{An Estimate of the Chemical Composition of Titan's
  Lakes}}.
\newblock \bibinfo{journal}{Astrophysical Journal letters}
  \bibinfo{volume}{707}, \bibinfo{pages}{L128--L131}.
\newblock \DOIprefix\doi{10.1088/0004-637X/707/2/L128},
  \href{http://arxiv.org/abs/0911.1860}{\tt arXiv:0911.1860}.
\bibitem[{{Cours} et~al.(2020){Cours}, {Cordier}, {Seignovert}, {Maltagliati}
  and {Biennier}}]{Cours2020}
\bibinfo{author}{{Cours}, T.}, \bibinfo{author}{{Cordier}, D.},
  \bibinfo{author}{{Seignovert}, B.}, \bibinfo{author}{{Maltagliati}, L.},
  \bibinfo{author}{{Biennier}, L.}, \bibinfo{year}{2020}.
\newblock \bibinfo{title}{{The 3 . 4 {\ensuremath{\mu}}m absorption in Titan's
  stratosphere: Contribution of ethane, propane, butane and complex
  hydrogenated organics}}.
\newblock \bibinfo{journal}{Icarus} \bibinfo{volume}{339},
  \bibinfo{pages}{113571}.
\newblock \DOIprefix\doi{10.1016/j.icarus.2019.113571},
  \href{http://arxiv.org/abs/2001.02791}{\tt arXiv:2001.02791}.
\bibitem[{{Crovisier} et~al.(2004){Crovisier}, {Bockel{\'e}e-Morvan}, {Colom},
  {Biver}, {Despois}, {Lis} and {Teamtarget-of-opportunity radio observations
  of comets}}]{Crovisier2004}
\bibinfo{author}{{Crovisier}, J.}, \bibinfo{author}{{Bockel{\'e}e-Morvan}, D.},
  \bibinfo{author}{{Colom}, P.}, \bibinfo{author}{{Biver}, N.},
  \bibinfo{author}{{Despois}, D.}, \bibinfo{author}{{Lis}, D.C.},
  \bibinfo{author}{{Teamtarget-of-opportunity radio observations of comets}},
  \bibinfo{year}{2004}.
\newblock \bibinfo{title}{{The composition of ices in comet C/1995 O1
  (Hale-Bopp) from radio spectroscopy. Further results and upper limits on
  undetected species}}.
\newblock \bibinfo{journal}{Astronomy and Astrophysics} \bibinfo{volume}{418},
  \bibinfo{pages}{1141--1157}.
\newblock \DOIprefix\doi{10.1051/0004-6361:20035688}.
\bibitem[{{Dartois} and {Deboffle}(2008)}]{Dartois2008}
\bibinfo{author}{{Dartois}, E.}, \bibinfo{author}{{Deboffle}, D.},
  \bibinfo{year}{2008}.
\newblock \bibinfo{title}{{Methane clathrate hydrate FTIR spectrum.
  Implications for its cometary and planetary detection}}.
\newblock \bibinfo{journal}{Astronomy and Astrophysics} \bibinfo{volume}{490},
  \bibinfo{pages}{L19--L22}.
\newblock \DOIprefix\doi{10.1051/0004-6361:200810926}.
\bibitem[{{Dartois} et~al.(2010){Dartois}, {Deboffle} and
  {Bouzit}}]{Dartois2010}
\bibinfo{author}{{Dartois}, E.}, \bibinfo{author}{{Deboffle}, D.},
  \bibinfo{author}{{Bouzit}, M.}, \bibinfo{year}{2010}.
\newblock \bibinfo{title}{{Methane clathrate hydrate infrared spectrum. II.
  Near-infrared overtones, combination modes and cages assignments}}.
\newblock \bibinfo{journal}{Astronomy and Astrophysics} \bibinfo{volume}{514},
  \bibinfo{pages}{A49}.
\newblock \DOIprefix\doi{10.1051/0004-6361/200913642}.
\bibitem[{{Dartois} and {Schmitt}(2009)}]{Dartois2009}
\bibinfo{author}{{Dartois}, E.}, \bibinfo{author}{{Schmitt}, B.},
  \bibinfo{year}{2009}.
\newblock \bibinfo{title}{{Carbon dioxide clathrate hydrate FTIR spectrum. Near
  infrared combination modes for astrophysical remote detection}}.
\newblock \bibinfo{journal}{Astronomy and Astrophysics} \bibinfo{volume}{504},
  \bibinfo{pages}{869--873}.
\newblock \DOIprefix\doi{10.1051/0004-6361/200911812}.
\bibitem[{{DeMeo} et~al.(2010){DeMeo}, {Dumas}, {de Bergh}, {Protopapa},
  {Cruikshank}, {Geballe}, {Alvarez-Candal}, {Merlin} and
  {Barucci}}]{DeMeo2010}
\bibinfo{author}{{DeMeo}, F.E.}, \bibinfo{author}{{Dumas}, C.},
  \bibinfo{author}{{de Bergh}, C.}, \bibinfo{author}{{Protopapa}, S.},
  \bibinfo{author}{{Cruikshank}, D.P.}, \bibinfo{author}{{Geballe}, T.R.},
  \bibinfo{author}{{Alvarez-Candal}, A.}, \bibinfo{author}{{Merlin}, F.},
  \bibinfo{author}{{Barucci}, M.A.}, \bibinfo{year}{2010}.
\newblock \bibinfo{title}{{A search for ethane on Pluto and Triton}}.
\newblock \bibinfo{journal}{Icarus} \bibinfo{volume}{208},
  \bibinfo{pages}{412--424}.
\newblock \DOIprefix\doi{10.1016/j.icarus.2010.01.014}.
\bibitem[{{Falenty} et~al.(2013){Falenty}, {Salamantin} and
  {Kuhs}}]{Falenty2013}
\bibinfo{author}{{Falenty}, A.}, \bibinfo{author}{{Salamantin}, A.N.},
  \bibinfo{author}{{Kuhs}, W.F.}, \bibinfo{year}{2013}.
\newblock \bibinfo{title}{{Kinetics of CO2‐Hydrate Formation from Ice
  Powders: Data Summary and Modeling Extended to Low Temperatures}}.
\newblock \bibinfo{journal}{J. Phys. Chem. C} \bibinfo{volume}{117},
  \bibinfo{pages}{8443--8457}.
\newblock \DOIprefix\doi{10.1021/jp310972b}.
\bibitem[{{Farnsworth} et~al.(2019){Farnsworth}, {Soderblom}, {Rodriguez},
  {Czaplinski} and {Chevrier}}]{Farnsworth2019}
\bibinfo{author}{{Farnsworth}, K.}, \bibinfo{author}{{Soderblom}, J.},
  \bibinfo{author}{{Rodriguez}, S.}, \bibinfo{author}{{Czaplinski}, E.},
  \bibinfo{author}{{Chevrier}, V.}, \bibinfo{year}{2019}.
\newblock \bibinfo{title}{{Constraining Ethane Concentration in Titan's Lakes
  and Seas.}}, in: \bibinfo{booktitle}{EPSC-DPS Joint Meeting 2019}, pp.
  \bibinfo{pages}{EPSC--DPS2019--1220}.
\bibitem[{{Fu} et~al.(2017){Fu}, {Ermakov}, {Marchi}, {Castillo-Rogez},
  {Raymond}, {Hager}, {Zuber}, {King}, {Bland }, {Cristina De Sanctis},
  {Preusker}, {Park} and {Russell}}]{Fu2017}
\bibinfo{author}{{Fu}, R.R.}, \bibinfo{author}{{Ermakov}, A.I.},
  \bibinfo{author}{{Marchi}, S.}, \bibinfo{author}{{Castillo-Rogez}, J.C.},
  \bibinfo{author}{{Raymond}, C.A.}, \bibinfo{author}{{Hager}, B.H.},
  \bibinfo{author}{{Zuber}, M.T.}, \bibinfo{author}{{King}, S.D.},
  \bibinfo{author}{{Bland }, M.T.}, \bibinfo{author}{{Cristina De Sanctis},
  M.}, \bibinfo{author}{{Preusker}, F.}, \bibinfo{author}{{Park}, R.S.},
  \bibinfo{author}{{Russell}, C.T.}, \bibinfo{year}{2017}.
\newblock \bibinfo{title}{{The interior structure of Ceres as revealed by
  surface topography}}.
\newblock \bibinfo{journal}{Earth and Planetary Science Letters}
  \bibinfo{volume}{476}, \bibinfo{pages}{153--164}.
\newblock \DOIprefix\doi{10.1016/j.epsl.2017.07.053}.
\bibitem[{Glein and Zolotov(2020)}]{Glein2020}
\bibinfo{author}{Glein, C.R.}, \bibinfo{author}{Zolotov, M.Y.},
  \bibinfo{year}{2020}.
\newblock \bibinfo{title}{Hydrogen, hydrocarbons, and habitability across the
  solar system}.
\newblock \bibinfo{journal}{Elements: An International Magazine of Mineralogy,
  Geochemistry, and Petrology} \bibinfo{volume}{16}, \bibinfo{pages}{47--52}.
\bibitem[{{Griffith} et~al.(2006){Griffith}, {Penteado}, {Rannou}, {Brown},
  {Boudon}, {Baines}, {Clark}, {Drossart}, {Buratti}, {Nicholson}, {McKay},
  {Coustenis}, {Negrao} and {Jaumann}}]{Griffith2006}
\bibinfo{author}{{Griffith}, C.A.}, \bibinfo{author}{{Penteado}, P.},
  \bibinfo{author}{{Rannou}, P.}, \bibinfo{author}{{Brown}, R.},
  \bibinfo{author}{{Boudon}, V.}, \bibinfo{author}{{Baines}, K.H.},
  \bibinfo{author}{{Clark}, R.}, \bibinfo{author}{{Drossart}, P.},
  \bibinfo{author}{{Buratti}, B.}, \bibinfo{author}{{Nicholson}, P.},
  \bibinfo{author}{{McKay}, C.P.}, \bibinfo{author}{{Coustenis}, A.},
  \bibinfo{author}{{Negrao}, A.}, \bibinfo{author}{{Jaumann}, R.},
  \bibinfo{year}{2006}.
\newblock \bibinfo{title}{{Evidence for a Polar Ethane Cloud on Titan}}.
\newblock \bibinfo{journal}{Science} \bibinfo{volume}{313},
  \bibinfo{pages}{1620--1622}.
\newblock \DOIprefix\doi{10.1126/science.1128245}.
\bibitem[{{Griffith} et~al.(2019){Griffith}, {Penteado}, {Turner}, {Neish},
  {Mitri}, {Montiel}, {Schoenfeld} and {Lopes}}]{Griffith2019}
\bibinfo{author}{{Griffith}, C.A.}, \bibinfo{author}{{Penteado}, P.F.},
  \bibinfo{author}{{Turner}, J.D.}, \bibinfo{author}{{Neish}, C.D.},
  \bibinfo{author}{{Mitri}, G.}, \bibinfo{author}{{Montiel}, N.J.},
  \bibinfo{author}{{Schoenfeld}, A.}, \bibinfo{author}{{Lopes}, R.M.C.},
  \bibinfo{year}{2019}.
\newblock \bibinfo{title}{{A corridor of exposed ice-rich bedrock across
  Titan's tropical region}}.
\newblock \bibinfo{journal}{Nature Astronomy} \bibinfo{volume}{3},
  \bibinfo{pages}{642--648}.
\newblock \DOIprefix\doi{10.1038/s41550-019-0756-5}.
\bibitem[{{Guerlet} et~al.(2009){Guerlet}, {Fouchet}, {B{\'e}zard},
  {Simon-Miller} and {Michael Flasar}}]{Guerlet2009}
\bibinfo{author}{{Guerlet}, S.}, \bibinfo{author}{{Fouchet}, T.},
  \bibinfo{author}{{B{\'e}zard}, B.}, \bibinfo{author}{{Simon-Miller}, A.A.},
  \bibinfo{author}{{Michael Flasar}, F.}, \bibinfo{year}{2009}.
\newblock \bibinfo{title}{{Vertical and meridional distribution of ethane,
  acetylene and propane in Saturn{\textquoteright}s stratosphere from
  CIRS/Cassini limb observations}}.
\newblock \bibinfo{journal}{Icarus} \bibinfo{volume}{203},
  \bibinfo{pages}{214--232}.
\newblock \DOIprefix\doi{10.1016/j.icarus.2009.04.002}.
\bibitem[{Guzm{\'a}n-Marmolejo and Segura(2015)}]{Guzman2015}
\bibinfo{author}{Guzm{\'a}n-Marmolejo, A.}, \bibinfo{author}{Segura, A.},
  \bibinfo{year}{2015}.
\newblock \bibinfo{title}{Methane in the solar system}.
\newblock \bibinfo{journal}{Bolet{\'\i}n de la Sociedad Geol{\'o}gica Mexicana}
  \bibinfo{volume}{67}, \bibinfo{pages}{377--385}.
\bibitem[{Hanley(2017)}]{Hanley2017}
\bibinfo{author}{Hanley, J.}, \bibinfo{year}{2017}.
\newblock \bibinfo{title}{Titan: Bubbles in focus}.
\newblock \bibinfo{journal}{Nature Astronomy} \bibinfo{volume}{1},
  \bibinfo{pages}{1--2}.
\bibitem[{{Hayes} et~al.(2008){Hayes}, {Aharonson}, {Callahan}, {Elachi},
  {Gim}, {Kirk}, {Lewis}, {Lopes}, {Lorenz}, {Lunine}, {Mitchell}, {Mitri},
  {Stofan} and {Wall}}]{Hayes2008}
\bibinfo{author}{{Hayes}, A.}, \bibinfo{author}{{Aharonson}, O.},
  \bibinfo{author}{{Callahan}, P.}, \bibinfo{author}{{Elachi}, C.},
  \bibinfo{author}{{Gim}, Y.}, \bibinfo{author}{{Kirk}, R.},
  \bibinfo{author}{{Lewis}, K.}, \bibinfo{author}{{Lopes}, R.},
  \bibinfo{author}{{Lorenz}, R.}, \bibinfo{author}{{Lunine}, J.},
  \bibinfo{author}{{Mitchell}, K.}, \bibinfo{author}{{Mitri}, G.},
  \bibinfo{author}{{Stofan}, E.}, \bibinfo{author}{{Wall}, S.},
  \bibinfo{year}{2008}.
\newblock \bibinfo{title}{{Hydrocarbon lakes on Titan: Distribution and
  interaction with a porous regolith}}.
\newblock \bibinfo{journal}{Geophysics Research Letters} \bibinfo{volume}{35},
  \bibinfo{pages}{L09204}.
\newblock \DOIprefix\doi{10.1029/2008GL033409}.
\bibitem[{{Hepp} et~al.(1997){Hepp}, {Georges} and {Herman}}]{Hepp1997}
\bibinfo{author}{{Hepp}, M.}, \bibinfo{author}{{Georges}, R.},
  \bibinfo{author}{{Herman}, M.}, \bibinfo{year}{1997}.
\newblock \bibinfo{title}{{The v$_{6}$ + v$_{10}$ band of ethane}}.
\newblock \bibinfo{journal}{Chemical Physics Letters} \bibinfo{volume}{275},
  \bibinfo{pages}{513--518}.
\newblock \DOIprefix\doi{10.1016/S0009-2614(97)00778-1}.
\bibitem[{{Hepp} and {Herman}(1998)}]{Hepp1998}
\bibinfo{author}{{Hepp}, M.}, \bibinfo{author}{{Herman}, M.},
  \bibinfo{year}{1998}.
\newblock \bibinfo{title}{{The jet cooled spectrum of ethane between 4000 and
  4500cm-1}}.
\newblock \bibinfo{journal}{Molecular Physics} \bibinfo{volume}{94},
  \bibinfo{pages}{829--838}.
\newblock \DOIprefix\doi{10.1080/00268979809482376}.
\bibitem[{{Hepp} and {Herman}(1999a)}]{Hepp1999b}
\bibinfo{author}{{Hepp}, M.}, \bibinfo{author}{{Herman}, M.},
  \bibinfo{year}{1999}a.
\newblock \bibinfo{title}{{Effective Rotation-Vibration Parameters for the
  {\ensuremath{\nu}} $_{8}$and {\ensuremath{\nu}} $_{4}$+ {\ensuremath{\nu}}
  $_{12}$Bands of Ethane}}.
\newblock \bibinfo{journal}{Journal of Molecular Spectroscopy}
  \bibinfo{volume}{194}, \bibinfo{pages}{87--94}.
\newblock \DOIprefix\doi{10.1006/jmsp.1998.7772}.
\bibitem[{{Hepp} and {Herman}(1999b)}]{Hepp1999}
\bibinfo{author}{{Hepp}, M.}, \bibinfo{author}{{Herman}, M.},
  \bibinfo{year}{1999}b.
\newblock \bibinfo{title}{{Weak Combination Bands in the 3-{\ensuremath{\mu}}m
  Region of Ethane}}.
\newblock \bibinfo{journal}{Journal of Molecular Spectroscopy}
  \bibinfo{volume}{197}, \bibinfo{pages}{56--63}.
\newblock \DOIprefix\doi{10.1006/jmsp.1999.7893}.
\bibitem[{{Hepp} and {Herman}(2000)}]{Hepp2000}
\bibinfo{author}{{Hepp}, M.}, \bibinfo{author}{{Herman}, M.},
  \bibinfo{year}{2000}.
\newblock \bibinfo{title}{{RESEARCH NOTE Vibration-rotation bands in ethane}}.
\newblock \bibinfo{journal}{Molecular Physics} \bibinfo{volume}{98},
  \bibinfo{pages}{57--61}.
\newblock \DOIprefix\doi{10.1080/00268970009483269}.
\bibitem[{{H{\"o}rst}(2017)}]{Horst2017}
\bibinfo{author}{{H{\"o}rst}, S.M.}, \bibinfo{year}{2017}.
\newblock \bibinfo{title}{{Titan's atmosphere and climate}}.
\newblock \bibinfo{journal}{Journal of Geophysical Research (Planets)}
  \bibinfo{volume}{122}, \bibinfo{pages}{432--482}.
\newblock \DOIprefix\doi{10.1002/2016JE005240},
  \href{http://arxiv.org/abs/1702.08611}{\tt arXiv:1702.08611}.
\bibitem[{{Hudson} et~al.(2014){Hudson}, {Gerakines} and {Moore}}]{Hudson2014}
\bibinfo{author}{{Hudson}, R.L.}, \bibinfo{author}{{Gerakines}, P.A.},
  \bibinfo{author}{{Moore}, M.H.}, \bibinfo{year}{2014}.
\newblock \bibinfo{title}{{Infrared spectra and optical constants of
  astronomical ices: II. Ethane and ethylene}}.
\newblock \bibinfo{journal}{Icarus} \bibinfo{volume}{243},
  \bibinfo{pages}{148--157}.
\newblock \DOIprefix\doi{10.1016/j.icarus.2014.09.001}.
\bibitem[{{Kalousov{\'a}} and {Sotin}(2020)}]{Kalousova2020}
\bibinfo{author}{{Kalousov{\'a}}, K.}, \bibinfo{author}{{Sotin}, C.},
  \bibinfo{year}{2020}.
\newblock \bibinfo{title}{{The Insulating Effect of Methane Clathrate Crust on
  Titan's Thermal Evolution}}.
\newblock \bibinfo{journal}{Geophysical Research Letters} \bibinfo{volume}{47},
  \bibinfo{pages}{e87481}.
\newblock \DOIprefix\doi{10.1029/2020GL087481}.
\bibitem[{Kida et~al.(2010)Kida, Jin, Takahashi, Nagao and Narita}]{Kida2010}
\bibinfo{author}{Kida, M.}, \bibinfo{author}{Jin, Y.},
  \bibinfo{author}{Takahashi, N.}, \bibinfo{author}{Nagao, J.},
  \bibinfo{author}{Narita, H.}, \bibinfo{year}{2010}.
\newblock \bibinfo{title}{Dissociation behavior of methane- ethane mixed gas
  hydrate coexisting structures i and ii}.
\newblock \bibinfo{journal}{The Journal of Physical Chemistry A}
  \bibinfo{volume}{114}, \bibinfo{pages}{9456--9461}.
\bibitem[{{Lombardo} et~al.(2019){Lombardo}, {Nixon}, {Sylvestre}, {Jennings},
  {Teanby}, {Irwin} and {Flasar}}]{Lombardo2019}
\bibinfo{author}{{Lombardo}, N.A.}, \bibinfo{author}{{Nixon}, C.A.},
  \bibinfo{author}{{Sylvestre}, M.}, \bibinfo{author}{{Jennings}, D.E.},
  \bibinfo{author}{{Teanby}, N.}, \bibinfo{author}{{Irwin}, P.J.G.},
  \bibinfo{author}{{Flasar}, F.M.}, \bibinfo{year}{2019}.
\newblock \bibinfo{title}{{Ethane in Titan{\textquoteright}s Stratosphere from
  Cassini CIRS Far- and Mid-infrared Spectra}}.
\newblock \bibinfo{journal}{Astronomical Journal} \bibinfo{volume}{157},
  \bibinfo{pages}{160}.
\newblock \DOIprefix\doi{10.3847/1538-3881/ab0e07},
  \href{http://arxiv.org/abs/1908.01926}{\tt arXiv:1908.01926}.
\bibitem[{Lorenz et~al.(2018)Lorenz, Turtle, Barnes, Trainer, Adams, Hibbard,
  Sheldon, Zacny, Peplowski, Lawrence et~al.}]{Lorenz2018}
\bibinfo{author}{Lorenz, R.D.}, \bibinfo{author}{Turtle, E.P.},
  \bibinfo{author}{Barnes, J.W.}, \bibinfo{author}{Trainer, M.G.},
  \bibinfo{author}{Adams, D.S.}, \bibinfo{author}{Hibbard, K.E.},
  \bibinfo{author}{Sheldon, C.Z.}, \bibinfo{author}{Zacny, K.},
  \bibinfo{author}{Peplowski, P.N.}, \bibinfo{author}{Lawrence, D.J.}, et~al.,
  \bibinfo{year}{2018}.
\newblock \bibinfo{title}{Dragonfly: A rotorcraft lander concept for scientific
  exploration at titan}.
\newblock \bibinfo{journal}{Johns Hopkins APL Technical Digest}
  \bibinfo{volume}{34}, \bibinfo{pages}{14}.
\bibitem[{{Lunine} and {Atreya}(2008)}]{Lunine2008}
\bibinfo{author}{{Lunine}, J.I.}, \bibinfo{author}{{Atreya}, S.K.},
  \bibinfo{year}{2008}.
\newblock \bibinfo{title}{{The methane cycle on Titan}}.
\newblock \bibinfo{journal}{Nature Geoscience} \bibinfo{volume}{1},
  \bibinfo{pages}{159--164}.
\newblock \DOIprefix\doi{10.1038/ngeo125}.
\bibitem[{{Lunine} et~al.(1983){Lunine}, {Stevenson} and {Yung}}]{Lunine1983}
\bibinfo{author}{{Lunine}, J.I.}, \bibinfo{author}{{Stevenson}, D.J.},
  \bibinfo{author}{{Yung}, Y.L.}, \bibinfo{year}{1983}.
\newblock \bibinfo{title}{{Ethane Ocean on Titan}}.
\newblock \bibinfo{journal}{Science} \bibinfo{volume}{222},
  \bibinfo{pages}{1229--1230}.
\newblock \DOIprefix\doi{10.1126/science.222.4629.1229}.
\bibitem[{{Luspay-Kuti} et~al.(2016){Luspay-Kuti}, {Mousis}, {H{\"a}ssig},
  {Fuselier}, {Lunine}, {Marty}, {Mand t}, {Wurz} and
  {Rubin}}]{Luspay-Kuti2016}
\bibinfo{author}{{Luspay-Kuti}, A.}, \bibinfo{author}{{Mousis}, O.},
  \bibinfo{author}{{H{\"a}ssig}, M.}, \bibinfo{author}{{Fuselier}, S.A.},
  \bibinfo{author}{{Lunine}, J.I.}, \bibinfo{author}{{Marty}, B.},
  \bibinfo{author}{{Mand t}, K.E.}, \bibinfo{author}{{Wurz}, P.},
  \bibinfo{author}{{Rubin}, M.}, \bibinfo{year}{2016}.
\newblock \bibinfo{title}{{The presence of clathrates in comet
  67P/Churyumov-Gerasimenko}}.
\newblock \bibinfo{journal}{Science Advances} \bibinfo{volume}{2},
  \bibinfo{pages}{1501781}.
\newblock \DOIprefix\doi{10.1126/sciadv.1501781}.
\bibitem[{{Marboeuf} et~al.(2012){Marboeuf}, {Schmitt}, {Petit}, {Mousis} and
  {Fray}}]{Marboeuf2012}
\bibinfo{author}{{Marboeuf}, U.}, \bibinfo{author}{{Schmitt}, B.},
  \bibinfo{author}{{Petit}, J.M.}, \bibinfo{author}{{Mousis}, O.},
  \bibinfo{author}{{Fray}, N.}, \bibinfo{year}{2012}.
\newblock \bibinfo{title}{{A cometary nucleus model taking into account all
  phase changes of water ice: amorphous, crystalline, and clathrate}}.
\newblock \bibinfo{journal}{Astronomie and Astrophysics} \bibinfo{volume}{542},
  \bibinfo{pages}{A82}.
\newblock \DOIprefix\doi{10.1051/0004-6361/201118176}.
\bibitem[{{Marounina} et~al.(2018){Marounina}, {Grasset}, {Tobie} and
  {Carpy}}]{Marounina2018}
\bibinfo{author}{{Marounina}, N.}, \bibinfo{author}{{Grasset}, O.},
  \bibinfo{author}{{Tobie}, G.}, \bibinfo{author}{{Carpy}, S.},
  \bibinfo{year}{2018}.
\newblock \bibinfo{title}{{Role of the global water ocean on the evolution of
  Titan's primitive atmosphere}}.
\newblock \bibinfo{journal}{Icarus} \bibinfo{volume}{310},
  \bibinfo{pages}{127--139}.
\newblock \DOIprefix\doi{10.1016/j.icarus.2017.10.048},
  \href{http://arxiv.org/abs/1711.09128}{\tt arXiv:1711.09128}.
\bibitem[{{Mastrogiuseppe} et~al.(2019){Mastrogiuseppe}, {Poggiali}, {Hayes},
  {Lunine}, {Seu}, {Mitri} and {Lorenz}}]{Mastrogiuseppe2019}
\bibinfo{author}{{Mastrogiuseppe}, M.}, \bibinfo{author}{{Poggiali}, V.},
  \bibinfo{author}{{Hayes}, A.G.}, \bibinfo{author}{{Lunine}, J.I.},
  \bibinfo{author}{{Seu}, R.}, \bibinfo{author}{{Mitri}, G.},
  \bibinfo{author}{{Lorenz}, R.D.}, \bibinfo{year}{2019}.
\newblock \bibinfo{title}{{Deep and methane-rich lakes on Titan}}.
\newblock \bibinfo{journal}{Nature Astronomy} \bibinfo{volume}{3},
  \bibinfo{pages}{535--542}.
\newblock \DOIprefix\doi{10.1038/s41550-019-0714-2}.
\bibitem[{{McKay} et~al.(1989){McKay}, {Pollack} and {Courtin}}]{McKay1989}
\bibinfo{author}{{McKay}, C.P.}, \bibinfo{author}{{Pollack}, J.B.},
  \bibinfo{author}{{Courtin}, R.}, \bibinfo{year}{1989}.
\newblock \bibinfo{title}{{The thermal structure of Titan's atmosphere}}.
\newblock \bibinfo{journal}{Icarus} \bibinfo{volume}{80},
  \bibinfo{pages}{23--53}.
\newblock \DOIprefix\doi{10.1016/0019-1035(89)90160-7}.
\bibitem[{{Melin} et~al.(2020){Melin}, {Fletcher}, {Irwin} and
  {Edgington}}]{Melin2020}
\bibinfo{author}{{Melin}, H.}, \bibinfo{author}{{Fletcher}, L.N.},
  \bibinfo{author}{{Irwin}, P.G.J.}, \bibinfo{author}{{Edgington}, S.G.},
  \bibinfo{year}{2020}.
\newblock \bibinfo{title}{{Jupiter in the Ultraviolet: Acetylene and Ethane
  Abundances in the Stratosphere of Jupiter from Cassini Observations between
  0.15 and 0.19 {\ensuremath{\mu}}m}}.
\newblock \bibinfo{journal}{Astronomical Journal} \bibinfo{volume}{159},
  \bibinfo{pages}{291}.
\newblock \DOIprefix\doi{10.3847/1538-3881/ab91a6},
  \href{http://arxiv.org/abs/2005.09895}{\tt arXiv:2005.09895}.
\bibitem[{{Mousis} and {Schmitt}(2008)}]{Mousis2008}
\bibinfo{author}{{Mousis}, O.}, \bibinfo{author}{{Schmitt}, B.},
  \bibinfo{year}{2008}.
\newblock \bibinfo{title}{{Sequestration of Ethane in the Cryovolcanic
  Subsurface of Titan}}.
\newblock \bibinfo{journal}{Astrophysical Journal, Letters}
  \bibinfo{volume}{677}, \bibinfo{pages}{L67}.
\newblock \DOIprefix\doi{10.1086/587141},
  \href{http://arxiv.org/abs/0802.1033}{\tt arXiv:0802.1033}.
\bibitem[{{Mumma} et~al.(2000){Mumma}, {DiSanti}, {Dello Russo}, {Magee-Sauer}
  and {Rettig}}]{Mumma2000}
\bibinfo{author}{{Mumma}, M.J.}, \bibinfo{author}{{DiSanti}, M.A.},
  \bibinfo{author}{{Dello Russo}, N.}, \bibinfo{author}{{Magee-Sauer}, K.},
  \bibinfo{author}{{Rettig}, T.W.}, \bibinfo{year}{2000}.
\newblock \bibinfo{title}{{Detection of CO and Ethane in Comet
  21P/Giacobini-Zinner: Evidence for Variable Chemistry in the Outer Solar
  Nebula}}.
\newblock \bibinfo{journal}{Astrophysical Journal letters}
  \bibinfo{volume}{531}, \bibinfo{pages}{L155--L159}.
\newblock \DOIprefix\doi{10.1086/312530}.
\bibitem[{{Mumma} et~al.(2001){Mumma}, {McLean}, {DiSanti}, {Larkin}, {Dello
  Russo}, {Magee-Sauer}, {Becklin}, {Bida}, {Chaffee}, {Conrad}, {Figer},
  {Gilbert}, {Graham}, {Levenson}, {Novak}, {Reuter}, {Teplitz}, {Wilcox} and
  {Xu}}]{Mumma2001}
\bibinfo{author}{{Mumma}, M.J.}, \bibinfo{author}{{McLean}, I.S.},
  \bibinfo{author}{{DiSanti}, M.A.}, \bibinfo{author}{{Larkin}, J.E.},
  \bibinfo{author}{{Dello Russo}, N.}, \bibinfo{author}{{Magee-Sauer}, K.},
  \bibinfo{author}{{Becklin}, E.E.}, \bibinfo{author}{{Bida}, T.},
  \bibinfo{author}{{Chaffee}, F.}, \bibinfo{author}{{Conrad}, A.R.},
  \bibinfo{author}{{Figer}, D.F.}, \bibinfo{author}{{Gilbert}, A.M.},
  \bibinfo{author}{{Graham}, J.R.}, \bibinfo{author}{{Levenson}, N.A.},
  \bibinfo{author}{{Novak}, R.E.}, \bibinfo{author}{{Reuter}, D.C.},
  \bibinfo{author}{{Teplitz}, H.I.}, \bibinfo{author}{{Wilcox}, M.K.},
  \bibinfo{author}{{Xu}, L.H.}, \bibinfo{year}{2001}.
\newblock \bibinfo{title}{{A Survey of Organic Volatile Species in Comet C/1999
  H1 (Lee) Using NIRSPEC at the Keck Observatory}}.
\newblock \bibinfo{journal}{Astrophysical Journal} \bibinfo{volume}{546},
  \bibinfo{pages}{1183--1193}.
\newblock \DOIprefix\doi{10.1086/318314}.
\bibitem[{Murshed and Kuhs(2009)}]{Murshed2009}
\bibinfo{author}{Murshed, M.M.}, \bibinfo{author}{Kuhs, W.F.},
  \bibinfo{year}{2009}.
\newblock \bibinfo{title}{Kinetic studies of methane--ethane mixed gas hydrates
  by neutron diffraction and raman spectroscopy}.
\newblock \bibinfo{journal}{The Journal of Physical Chemistry B}
  \bibinfo{volume}{113}, \bibinfo{pages}{5172--5180}.
\bibitem[{{Niemann} et~al.(2010){Niemann}, {Atreya}, {Demick}, {Gautier},
  {Haberman}, {Harpold}, {Kasprzak}, {Lunine}, {Owen} and
  {Raulin}}]{Niemann2010}
\bibinfo{author}{{Niemann}, H.B.}, \bibinfo{author}{{Atreya}, S.K.},
  \bibinfo{author}{{Demick}, J.E.}, \bibinfo{author}{{Gautier}, D.},
  \bibinfo{author}{{Haberman}, J.A.}, \bibinfo{author}{{Harpold}, D.N.},
  \bibinfo{author}{{Kasprzak}, W.T.}, \bibinfo{author}{{Lunine}, J.I.},
  \bibinfo{author}{{Owen}, T.C.}, \bibinfo{author}{{Raulin}, F.},
  \bibinfo{year}{2010}.
\newblock \bibinfo{title}{{Composition of Titan's lower atmosphere and simple
  surface volatiles as measured by the Cassini-Huygens probe gas chromatograph
  mass spectrometer experiment}}.
\newblock \bibinfo{journal}{Journal of Geophysical Research (Planets)}
  \bibinfo{volume}{115}, \bibinfo{pages}{E12006}.
\newblock \DOIprefix\doi{10.1029/2010JE003659}.
\bibitem[{Ohno et~al.(2009)Ohno, Strobel, Dec, Sloan and Koh}]{Ohno2009}
\bibinfo{author}{Ohno, H.}, \bibinfo{author}{Strobel, T.A.},
  \bibinfo{author}{Dec, S.F.}, \bibinfo{author}{Sloan, Jr, E.D.},
  \bibinfo{author}{Koh, C.A.}, \bibinfo{year}{2009}.
\newblock \bibinfo{title}{Raman studies of methane- ethane hydrate
  metastability}.
\newblock \bibinfo{journal}{The Journal of Physical Chemistry A}
  \bibinfo{volume}{113}, \bibinfo{pages}{1711--1716}.
\bibitem[{{Osegovic} and {Max}(2005)}]{Osegovic2005}
\bibinfo{author}{{Osegovic}, J.P.}, \bibinfo{author}{{Max}, M.D.},
  \bibinfo{year}{2005}.
\newblock \bibinfo{title}{{Compound clathrate hydrate on Titan's surface}}.
\newblock \bibinfo{journal}{Journal of Geophysical Research (Planets)}
  \bibinfo{volume}{110}, \bibinfo{pages}{E08004}.
\newblock \DOIprefix\doi{10.1029/2005JE002435}.
\bibitem[{{Richardson} et~al.(1985){Richardson}, {Woolridge} and
  {Devlin}}]{Richardson1985}
\bibinfo{author}{{Richardson}, H.H.}, \bibinfo{author}{{Woolridge}, P.J.},
  \bibinfo{author}{{Devlin}, J.P.}, \bibinfo{year}{1985}.
\newblock \bibinfo{title}{{FT-IR spectra of vacuum deposited clathrate hydrate
  of oxirane H\_2S, THF, and ethane}}.
\newblock \bibinfo{journal}{Journal of Chemical Physics} \bibinfo{volume}{83},
  \bibinfo{pages}{4387--4394}.
\newblock \DOIprefix\doi{10.1063/1.449055}.
\bibitem[{{Roberts} et~al.(1940){Roberts}, {Brownscombe}, {Howe} and
  {Ramser}}]{Roberts1940}
\bibinfo{author}{{Roberts}, O.}, \bibinfo{author}{{Brownscombe}, E.},
  \bibinfo{author}{{Howe}, L.}, \bibinfo{author}{{Ramser}, H.},
  \bibinfo{year}{1940}.
\newblock \bibinfo{title}{{Constitution diagrams and composition of methane and
  ethane hydrates}}.
\newblock \bibinfo{journal}{Oil \& Gas J.} \bibinfo{volume}{39},
  \bibinfo{pages}{37}.
\bibitem[{{Sada} et~al.(1996){Sada}, {McCabe}, {Bjoraker}, {Jennings} and
  {Reuter}}]{Sada1996}
\bibinfo{author}{{Sada}, P.V.}, \bibinfo{author}{{McCabe}, G.H.},
  \bibinfo{author}{{Bjoraker}, G.L.}, \bibinfo{author}{{Jennings}, D.E.},
  \bibinfo{author}{{Reuter}, D.C.}, \bibinfo{year}{1996}.
\newblock \bibinfo{title}{{13C-Ethane in the Atmospheres of Jupiter and
  Saturn}}.
\newblock \bibinfo{journal}{Astrophysical Journal} \bibinfo{volume}{472},
  \bibinfo{pages}{903}.
\newblock \DOIprefix\doi{10.1086/178120}.
\bibitem[{Sloan and Koh(2007)}]{Sloan2007}
\bibinfo{author}{Sloan, E.D.}, \bibinfo{author}{Koh, C.A.},
  \bibinfo{year}{2007}.
\newblock \bibinfo{title}{Clathrate hydrates of Natural Gases}.
\newblock \bibinfo{edition}{3} ed., \bibinfo{publisher}{CRC Press}.
\bibitem[{{Stofan} et~al.(2007){Stofan}, {Elachi}, {Lunine}, {Lorenz},
  {Stiles}, {Mitchell}, {Ostro}, {Soderblom}, {Wood}, {Zebker}, {Wall},
  {Janssen}, {Kirk}, {Lopes}, {Paganelli}, {Radebaugh}, {Wye}, {Anderson},
  {Allison}, {Boehmer}, {Callahan}, {Encrenaz}, {Flamini}, {Francescetti},
  {Gim}, {Hamilton}, {Hensley}, {Johnson}, {Kelleher}, {Muhleman}, {Paillou},
  {Picardi}, {Posa}, {Roth}, {Seu}, {Shaffer}, {Vetrella} and
  {West}}]{Stofan2007}
\bibinfo{author}{{Stofan}, E.R.}, \bibinfo{author}{{Elachi}, C.},
  \bibinfo{author}{{Lunine}, J.I.}, \bibinfo{author}{{Lorenz}, R.D.},
  \bibinfo{author}{{Stiles}, B.}, \bibinfo{author}{{Mitchell}, K.L.},
  \bibinfo{author}{{Ostro}, S.}, \bibinfo{author}{{Soderblom}, L.},
  \bibinfo{author}{{Wood}, C.}, \bibinfo{author}{{Zebker}, H.},
  \bibinfo{author}{{Wall}, S.}, \bibinfo{author}{{Janssen}, M.},
  \bibinfo{author}{{Kirk}, R.}, \bibinfo{author}{{Lopes}, R.},
  \bibinfo{author}{{Paganelli}, F.}, \bibinfo{author}{{Radebaugh}, J.},
  \bibinfo{author}{{Wye}, L.}, \bibinfo{author}{{Anderson}, Y.},
  \bibinfo{author}{{Allison}, M.}, \bibinfo{author}{{Boehmer}, R.},
  \bibinfo{author}{{Callahan}, P.}, \bibinfo{author}{{Encrenaz}, P.},
  \bibinfo{author}{{Flamini}, E.}, \bibinfo{author}{{Francescetti}, G.},
  \bibinfo{author}{{Gim}, Y.}, \bibinfo{author}{{Hamilton}, G.},
  \bibinfo{author}{{Hensley}, S.}, \bibinfo{author}{{Johnson}, W.T.K.},
  \bibinfo{author}{{Kelleher}, K.}, \bibinfo{author}{{Muhleman}, D.},
  \bibinfo{author}{{Paillou}, P.}, \bibinfo{author}{{Picardi}, G.},
  \bibinfo{author}{{Posa}, F.}, \bibinfo{author}{{Roth}, L.},
  \bibinfo{author}{{Seu}, R.}, \bibinfo{author}{{Shaffer}, S.},
  \bibinfo{author}{{Vetrella}, S.}, \bibinfo{author}{{West}, R.},
  \bibinfo{year}{2007}.
\newblock \bibinfo{title}{{The lakes of Titan}}.
\newblock \bibinfo{journal}{Nature} \bibinfo{volume}{445},
  \bibinfo{pages}{61--64}.
\newblock \DOIprefix\doi{10.1038/nature05438}.
\bibitem[{{Subramanian} et~al.(2000){Subramanian}, {Kini}, {Dec} and
  {Sloan}}]{Subramanian2000}
\bibinfo{author}{{Subramanian}, S.}, \bibinfo{author}{{Kini}, R.A.},
  \bibinfo{author}{{Dec}, S.F.}, \bibinfo{author}{{Sloan}, E.~D., J.},
  \bibinfo{year}{2000}.
\newblock \bibinfo{title}{{Structural Transition Studies in Methane + Ethane
  Hydrates Using Raman and NMR}}.
\newblock \bibinfo{journal}{Annals of the New York Academy of Sciences}
  \bibinfo{volume}{912}, \bibinfo{pages}{873--886}.
\newblock \DOIprefix\doi{10.1111/j.1749-6632.2000.tb06841.x}.
\bibitem[{{Takeya} et~al.(2003){Takeya}, {Kamata}, {Uchida}, {Nagao},
  {Ebinuma}, {Narita}, {Hori} and {Hondoh}}]{Takeya2003}
\bibinfo{author}{{Takeya}, S.}, \bibinfo{author}{{Kamata}, Y.},
  \bibinfo{author}{{Uchida}, T.}, \bibinfo{author}{{Nagao}, J.},
  \bibinfo{author}{{Ebinuma}, T.}, \bibinfo{author}{{Narita}, H.},
  \bibinfo{author}{{Hori}, A.}, \bibinfo{author}{{Hondoh}, T.},
  \bibinfo{year}{2003}.
\newblock \bibinfo{title}{{Coexistence of structure I and II hydrates formed
  from a mixture of methane and ethane gases}}.
\newblock \bibinfo{journal}{Canadian Journal of Physics} \bibinfo{volume}{81},
  \bibinfo{pages}{479--484}.
\newblock \DOIprefix\doi{10.1139/p03-038}.
\bibitem[{{Takeya} et~al.(2005){Takeya}, {Uchida}, {Kamata}, {Nagao}, {Kida},
  {Minami}, {Sakagami}, {Hachikubo}, {Takahashi}, {Shoji}, {Khlystov},
  {Grachev}, {Soloviev}, {Narita}, {Hori} and {Hondoh}}]{Takeya2005}
\bibinfo{author}{{Takeya}, S.}, \bibinfo{author}{{Uchida}, T.},
  \bibinfo{author}{{Kamata}, Y.}, \bibinfo{author}{{Nagao}, J.},
  \bibinfo{author}{{Kida}, M.}, \bibinfo{author}{{Minami}, H.},
  \bibinfo{author}{{Sakagami}, H.}, \bibinfo{author}{{Hachikubo}, A.},
  \bibinfo{author}{{Takahashi}, N.}, \bibinfo{author}{{Shoji}, H.},
  \bibinfo{author}{{Khlystov}, O.}, \bibinfo{author}{{Grachev}, M.},
  \bibinfo{author}{{Soloviev}, V.~{Ebinuma}, T.}, \bibinfo{author}{{Narita},
  H.}, \bibinfo{author}{{Hori}, A.}, \bibinfo{author}{{Hondoh}, T.},
  \bibinfo{year}{2005}.
\newblock \bibinfo{title}{{Lattice Expansion of Clathrate Hydrates of Methane
  Mixtures and Natural Gas}}.
\newblock \bibinfo{journal}{Angew. Chem.} \bibinfo{volume}{117},
  \bibinfo{pages}{7088 –7091}.
\newblock \DOIprefix\doi{10.1002/ange.200501845}.
\bibitem[{{Thomas} et~al.(2007){Thomas}, {Mousis}, {Ballenegger} and
  {Picaud}}]{Thomas2007}
\bibinfo{author}{{Thomas}, C.}, \bibinfo{author}{{Mousis}, O.},
  \bibinfo{author}{{Ballenegger}, V.}, \bibinfo{author}{{Picaud}, S.},
  \bibinfo{year}{2007}.
\newblock \bibinfo{title}{{Clathrate hydrates as a sink of noble gases in
  Titan's atmosphere}}.
\newblock \bibinfo{journal}{Astronomy and Astrophysics} \bibinfo{volume}{474},
  \bibinfo{pages}{L17--L20}.
\newblock \DOIprefix\doi{10.1051/0004-6361:20078072},
  \href{http://arxiv.org/abs/0708.2158}{\tt arXiv:0708.2158}.
\bibitem[{{Thomas} et~al.(2008){Thomas}, {Picaud}, {Mousis} and
  {Ballenegger}}]{Thomas2008}
\bibinfo{author}{{Thomas}, C.}, \bibinfo{author}{{Picaud}, S.},
  \bibinfo{author}{{Mousis}, O.}, \bibinfo{author}{{Ballenegger}, V.},
  \bibinfo{year}{2008}.
\newblock \bibinfo{title}{{A theoretical investigation into the trapping of
  noble gases by clathrates on Titan}}.
\newblock \bibinfo{journal}{Planetary and Space Science} \bibinfo{volume}{56},
  \bibinfo{pages}{1607--1617}.
\newblock \DOIprefix\doi{10.1016/j.pss.2008.04.009},
  \href{http://arxiv.org/abs/0803.2884}{\tt arXiv:0803.2884}.
\bibitem[{{Tobie} et~al.(2009){Tobie}, {Choukroun}, {Grasset}, {Le
  Mou{\'e}lic}, {Lunine}, {Sotin}, {Bourgeois}, {Gautier}, {Hirtzig},
  {Lebonnois} and {Le Corre}}]{Tobie2009}
\bibinfo{author}{{Tobie}, G.}, \bibinfo{author}{{Choukroun}, M.},
  \bibinfo{author}{{Grasset}, O.}, \bibinfo{author}{{Le Mou{\'e}lic}, S.},
  \bibinfo{author}{{Lunine}, J.I.}, \bibinfo{author}{{Sotin}, C.},
  \bibinfo{author}{{Bourgeois}, O.}, \bibinfo{author}{{Gautier}, D.},
  \bibinfo{author}{{Hirtzig}, M.}, \bibinfo{author}{{Lebonnois}, S.},
  \bibinfo{author}{{Le Corre}, L.}, \bibinfo{year}{2009}.
\newblock \bibinfo{title}{{Evolution of Titan and implications for its
  hydrocarbon cycle}}.
\newblock \bibinfo{journal}{Philosophical Transactions of the Royal Society of
  London Series A} \bibinfo{volume}{367}, \bibinfo{pages}{617--631}.
\newblock \DOIprefix\doi{10.1098/rsta.2008.0246}.
\bibitem[{{Tokunaga} et~al.(1975){Tokunaga}, {Knacke} and
  {Owen}}]{Tokunaga1975}
\bibinfo{author}{{Tokunaga}, A.}, \bibinfo{author}{{Knacke}, R.F.},
  \bibinfo{author}{{Owen}, T.}, \bibinfo{year}{1975}.
\newblock \bibinfo{title}{{The detection of ethane on Saturn.}}
\newblock \bibinfo{journal}{Astrophysical Journal, Letters}
  \bibinfo{volume}{197}, \bibinfo{pages}{L77}.
\newblock \DOIprefix\doi{10.1086/181782}.
\bibitem[{{Turtle} et~al.(2020){Turtle}, {Trainer}, {Barnes}, {Lorenz},
  {Hibbard}, {Adams}, {Bedini}, {Brinckerhoff}, {Burks}, {Cable}, {Ernst},
  {Freissinet}, {Hand}, {Hayes}, {H{\"o}rst}, {Johnson}, {Karkoschka},
  {Langelaan}, {Lawrence}, {Le Gall}, {Lora}, {MacKenzie}, {McKay}, {Miller},
  {Murchie}, {Neish}, {Newman}, {N{\'u}{\~n}ez}, {Palacios}, {Panning},
  {Parsons}, {Peplowski}, {Quick}, {Radebaugh}, {Rafkin}, {Ravine}, {Schmitz},
  {Shiraishi}, {Soderblom}, {Sotzen}, {Stickle}, {Stofan}, {Szopa}, {Tokano},
  {Wilson}, {Yingst} and {Zacny}}]{Turtle2020}
\bibinfo{author}{{Turtle}, E.P.}, \bibinfo{author}{{Trainer}, M.G.},
  \bibinfo{author}{{Barnes}, J.W.}, \bibinfo{author}{{Lorenz}, R.D.},
  \bibinfo{author}{{Hibbard}, K.E.}, \bibinfo{author}{{Adams}, D.S.},
  \bibinfo{author}{{Bedini}, P.D.}, \bibinfo{author}{{Brinckerhoff}, W.B.},
  \bibinfo{author}{{Burks}, M.T.}, \bibinfo{author}{{Cable}, M.L.},
  \bibinfo{author}{{Ernst}, C.}, \bibinfo{author}{{Freissinet}, C.},
  \bibinfo{author}{{Hand}, K.}, \bibinfo{author}{{Hayes}, A.G.},
  \bibinfo{author}{{H{\"o}rst}, S.M.}, \bibinfo{author}{{Johnson}, J.R.},
  \bibinfo{author}{{Karkoschka}, E.}, \bibinfo{author}{{Langelaan}, J.W.},
  \bibinfo{author}{{Lawrence}, D.J.}, \bibinfo{author}{{Le Gall}, A.},
  \bibinfo{author}{{Lora}, J.M.}, \bibinfo{author}{{MacKenzie}, S.M.},
  \bibinfo{author}{{McKay}, C.P.}, \bibinfo{author}{{Miller}, R.S.},
  \bibinfo{author}{{Murchie}, S.}, \bibinfo{author}{{Neish}, C.D.},
  \bibinfo{author}{{Newman}, C.E.}, \bibinfo{author}{{N{\'u}{\~n}ez}, J.I.},
  \bibinfo{author}{{Palacios}, J.}, \bibinfo{author}{{Panning}, M.P.},
  \bibinfo{author}{{Parsons}, A.M.}, \bibinfo{author}{{Peplowski}, P.N.},
  \bibinfo{author}{{Quick}, L.C.}, \bibinfo{author}{{Radebaugh}, J.},
  \bibinfo{author}{{Rafkin}, S.C.R.}, \bibinfo{author}{{Ravine}, M.A.},
  \bibinfo{author}{{Schmitz}, S.}, \bibinfo{author}{{Shiraishi}, H.},
  \bibinfo{author}{{Soderblom}, J.M.}, \bibinfo{author}{{Sotzen}, K.S.},
  \bibinfo{author}{{Stickle}, A.M.}, \bibinfo{author}{{Stofan}, E.R.},
  \bibinfo{author}{{Szopa}, C.}, \bibinfo{author}{{Tokano}, T.},
  \bibinfo{author}{{Wilson}, C.}, \bibinfo{author}{{Yingst}, R.A.},
  \bibinfo{author}{{Zacny}, K.}, \bibinfo{year}{2020}.
\newblock \bibinfo{title}{{Dragonfly: In Situ Exploration of Titan's Organic
  Chemistry and Habitability}}, in: \bibinfo{booktitle}{Lunar and Planetary
  Science Conference}, p. \bibinfo{pages}{2288}.
\bibitem[{{Villanueva} et~al.(2011){Villanueva}, {Mumma} and
  {Magee-Sauer}}]{Villanueva2011}
\bibinfo{author}{{Villanueva}, G.L.}, \bibinfo{author}{{Mumma}, M.J.},
  \bibinfo{author}{{Magee-Sauer}, K.}, \bibinfo{year}{2011}.
\newblock \bibinfo{title}{{Ethane in planetary and cometary atmospheres:
  Transmittance and fluorescence models of the {\ensuremath{\nu}}$_{7}$ band at
  3.3 {\ensuremath{\mu}}m}}.
\newblock \bibinfo{journal}{Journal of Geophysical Research (Planets)}
  \bibinfo{volume}{116}, \bibinfo{pages}{E08012}.
\newblock \DOIprefix\doi{10.1029/2010JE003794}.
\bibitem[{{Voosen}(2019)}]{Voosen2019}
\bibinfo{author}{{Voosen}, P.}, \bibinfo{year}{2019}.
\newblock \bibinfo{title}{{NASA to fly drone on Titan}}.
\newblock \bibinfo{journal}{Science} \bibinfo{volume}{365},
  \bibinfo{pages}{15--15}.
\newblock \DOIprefix\doi{10.1126/science.365.6448.15-a}.
\bibitem[{{Vu} et~al.(2020){Vu}, {Choukroun}, {Sotin}, {Mu{\~n}oz-Iglesias} and
  {Maynard-Casely}}]{Vu2020}
\bibinfo{author}{{Vu}, T.{\^A}.H.}, \bibinfo{author}{{Choukroun}, M.},
  \bibinfo{author}{{Sotin}, C.}, \bibinfo{author}{{Mu{\~n}oz-Iglesias}, V.},
  \bibinfo{author}{{Maynard-Casely}, H.{\^A}.E.}, \bibinfo{year}{2020}.
\newblock \bibinfo{title}{{Rapid Formation of Clathrate Hydrate From Liquid
  Ethane and Water Ice on Titan}}.
\newblock \bibinfo{journal}{Geophysics Research Letters} \bibinfo{volume}{47},
  \bibinfo{pages}{e86265}.
\newblock \DOIprefix\doi{10.1029/2019GL086265}.
\bibitem[{{Younglove} and {Ely}(1987)}]{Younglove1987}
\bibinfo{author}{{Younglove}, B.A.}, \bibinfo{author}{{Ely}, J.F.},
  \bibinfo{year}{1987}.
\newblock \bibinfo{title}{{Thermophysical Properties of Fluids. II. Methane,
  Ethane, Propane, Isobutane, and Normal Butane}}.
\newblock \bibinfo{journal}{Journal of Physical and Chemical Reference Data}
  \bibinfo{volume}{16}, \bibinfo{pages}{577--798}.
\newblock \DOIprefix\doi{10.1063/1.555785}.

\end{thebibliography}
\end{document}